\RequirePackage{ifpdf}
\documentclass[]{JHEP3}
\pdfoutput=1
\usepackage{amsmath}
\usepackage{amssymb}
\usepackage{epsfig,multicol,bbm}
\usepackage{graphicx}
\usepackage{bm}
\usepackage{multirow}
\usepackage{bbold}
\usepackage[mathscr]{euscript}

\renewcommand{\app}[1]{Appendix~\ref{#1}}
\newcommand{\Ref}[1]{Ref.~\cite{#1}}
\newcommand{\Refs}[2]{Refs.~\cite{#1,#2}}

\newcommand{\be}[0]{\begin{equation}}
\newcommand{\ee}[0]{\end{equation}}

\newcommand{\Bbar}{\bar B}
\newcommand{\yc}{y_{\rm cut}}
\newcommand{\pTc}{p_T^{\rm cut}}

\newcommand{\eq}[1]{Eq.~\eqref{eq:#1}}
\newcommand{\eqs}[2]{Eqs.~\eqref{eq:#1} and \eqref{eq:#2}}

\renewcommand{\sec}[1]{Sec.~\ref{sec:#1}}
\newcommand{\ssec}[1]{Sec.~\ref{ssec:#1}}

\newcommand{\tab}[1]{Table~\ref{tab:#1}}

\newcommand{\nn}{\nonumber}

\newcommand{\mcdot}{\!\cdot\!}

\newcommand{\vect}[1]{\mathbf{#1}}
\newcommand{\abs}[1]{\left\lvert #1\right\rvert}

\newcommand{\Ecm}{{\rm E_{cm}}}

\newcommand{\GeV}{\text{ GeV}}
\newcommand{\TeV}{\text{ TeV}}

\DeclareMathOperator{\Tr}{Tr}
\DeclareMathOperator{\Real}{Re}

\DeclareMathOperator{\Li}{Li}
\newcommand{\CF}{C_F}
\newcommand{\TR}{T_R}
\newcommand{\CA}{C_A}
\newcommand{\NF}{N_f}

\newcommand{\as}{\alpha_s}

\newcommand{\cO}{\mathcal{O}}
\newcommand{\cR}{\mathcal{R}}

\newcommand{\cI}{\mathcal{I}}
\newcommand{\cG}{\mathcal{G}}

\newcommand{\cJ}{\mathcal{J}}

\newcommand{\cusp}{c}

\newcommand{\meas}{{\rm meas}}
\newcommand{\unmeas}{ {\rm unmeas}}

\newcommand{\incl}{ {\rm incl}}

\newcommand{\tabrule}{\rule{-5pt}{3.5ex} \rule[-2.5ex]{0pt}{0pt}}
\newcommand{\cas}[1]{C_{#1}}
\newcommand{\pairs}{ i < j }
\newcommand{\hc}{{\rm h.c.}}
\newcommand{\GammaH}{\vect \Gamma _{\!  H}}

\renewcommand{\log}{\ln}

\newcommand{\Beam}{B}
\newcommand{\Beambar}{\bar B}

\newcommand{\pTJ}{p_T}

\newcommand{\Tij}{\vect{T}_i \mcdot \vect {T}_j}

\newcommand{\born}{\rm born}
\newcommand{\norm}{N}
\newcommand{\Norm}{ \frac{\pTJ }{8 \pi x_1 x_2 E^4_{\rm cm}}\frac{1}{\norm} }

\newcommand{\plus}{\!+}

\newcommand{\RH}{\vect{R}}

\newcommand{\PiH}{\vect{\Pi}_H}
\newcommand{\PiS}{\vect{\Pi}_S}

\newcommand{\muSunmeas}{\bar{\mu}_S}
\newcommand{\muJmeas}{\mu_J}
\newcommand{\muJunmeas}{\bar{\mu}_J}
\newcommand{\muSmeas}{\mu_S}

\newcommand{\mJunmeas}{\bar{m}_J}
\newcommand{\mJmeas}{m_J}

\newcommand{\gammaunmeas}{\Delta\gamma_{ss}}

\newcommand{\Ms}{\vect{M}'}
\newcommand{\Mh}{\vect{M}}

\title{Jet Shapes in Dijet Events at the LHC in SCET}

\author{
 Andrew Hornig \\
 Theoretical Division T-2, Los Alamos National Laboratory, Los Alamos, NM, 87545
 \\ E-mail: \email{ahornig@lanl.gov}
 }

\author{
 Yiannis Makris and Thomas Mehen \\
 Department of Physics, Duke University, Durham, NC 27708
 \\ E-mail: \email{yiannis.makris@duke.edu}, \email{mehen@phy.duke.edu}
 }

\preprint{
}	

\abstract{
We consider the class of jet shapes known as angularities in dijet production at hadron colliders. These angularities are modified from the original definitions in $e^+e^-$ collisions  to be boost invariant along the beam axis. These shapes apply to the constituents of jets defined with respect to either $k_T$-type (anti-$k_T$, $C/A$, and $k_T$) algorithms and cone-type algorithms. We present an SCET factorization formula and calculate the ingredients needed to achieve next-to-leading-log (NLL) accuracy in kinematic regions where non-global logarithms are not large. The factorization formula involves previously unstudied ``unmeasured beam functions,'' which are present for finite rapidity cuts around the beams.
We derive relations between the jet functions and the shape-dependent part of the soft function that appear in the factorized cross section and those previously calculated for $e^+e^-$ collisions, and present the calculation of the non-trivial, color-connected part of the soft-function to $\cO(\as)$. This latter part of the soft function is universal in the sense that it applies to any experimental setup with an out-of-jet $p_T$ veto and rapidity cuts together with two identified jets and it is independent of the choice of jet (sub-)structure measurement. In addition, we implement the recently introduced soft-collinear refactorization to resum logarithms of the jet size, valid in the region of non-enhanced non-global logarithm effects. While our results are valid for all $2 \to 2$ channels, we compute explicitly for the  $qq' \to qq'$ channel the color-flow matrices  and plot  the NLL resummed differential dijet cross section as  an explicit example, which shows that the normalization and scale uncertainty is reduced when the soft function is refactorized. For this channel, we also plot the jet size $R$ dependence, the $p_T^{\rm cut}$ dependence, and the dependence on the angularity parameter $a$. 
}

\preprint{LA-UR-15-27299}

\begin{document}



\section{Introduction}
\label{sec:intro}

Jet production is associated with a large number of important scattering processes at colliders such as the Large Hadron Collider (LHC). It is therefore crucial to have a robust  understanding of jets and jet production, and indeed much experimental and theoretical effort has gone into improving our understanding of jets. 
For hadron colliders, all theoretical predictions are based on the idea of QCD factorization \cite{Collins:1989gx,Sterman:1995fz}, which in its most basic form states that hadronic cross sections can be factorized into parton distribution functions (PDFs) and perturbatively calculable partonic cross sections. In multi-scale problems, these partonic cross sections can often be further factorized  into pieces which only depend on a single scale and the renormalization group evolution (RGE) of each piece from the single scale that it is sensitive to (its ``canonical scale'') to a common scale resums the logarithms of ratios of these scales which would otherwise spoil the perturbative convergence of the partonic cross section when   the scales are widely separated. An effective field theory approach to systematically factorizing cross sections is Soft-Collinear Effective Theory (SCET) \cite{Bauer:2000ew,Bauer:2000yr,Bauer:2001ct,Bauer:2001yt}.

A paradigmatic application of SCET is  the factorization and resummation of logarithms in event shapes measured in $e^+ e^-$ collisions \cite{Bauer:2002ie,Bauer:2003di,Fleming:2007qr,Bauer:2008dt}. Such event shapes, denoted by $e$, can often be defined so that they vanish in the limit of perfectly narrow  jets (so for example $e = 0$ for the tree-level process $e^+ e^- \to q\bar q$, and $e \to 0$ for events with additional radiation in the soft and collinear limits), and a fixed-order calculation of the cross section to $\cO(\as^n)$ would then contain logarithms of the form $(1/e)\alpha_s^n \ln^m e$ (for $m \leq 2n-1$). SCET factorization postulates that the partonic cross section can be written in terms a hard function $H$ which encapsulates the short-distance physics, jet functions $J$ that encapsulate collinear radiation within each jet and a soft function $S$ that encapsulates soft cross-talk between the jets, provided that the soft-collinear overlap (i.e. the `zero-bin') has been properly subtracted from the jet functions \cite{Manohar:2006nz}. For two back-to-back jets the factorization formula   takes the schematic form 
\begin{align}
d \sigma^{e^+ e^-} \sim H(Q) \times J_n(Q e^\alpha) \otimes J_{\bar n}(Q e^\alpha) \otimes S_{n \bar n}(Q e)
\,,\end{align}
where $\otimes$ denotes a convolution over $e$,    $n$ and $\bar n$ are the light-cone directions of the jets, the arguments of the functions denote the functions' canonical scales, $Q \sim \Ecm$ is a short-distance (hard) scale, and $\alpha$ is a parameter that depends on the choice of $e$ with $0<\alpha<1$ such that the canonical scales satisfy $Q e \ll Q e^\alpha \ll Q $ for $e \ll 1$. 
In the case of shapes which characterize multijet events (such as those of \cite{Ellis:2010rwa}), factorization simply involves more jet functions $J_{n_i}$ for each jet with direction $n_i$ and a more complicated soft function $S_{n_1 n_2 \cdots}$. 

One of the aims in the study of jet shapes is to study the internal energy patterns within a jet, i.e., the jet's substructure. This substructure can be used for example to help distinguish quark and gluon jets, or jets of purely QCD origin from those associated with other Standard Model mechanisms or from entirely new physics. Much work has recently been done on the analytical understanding of jet substructure, both for Monte Carlo event generator validation and for use as stand alone predictions \cite{Chien:2012ur,Dasgupta:2013ihk, Jouttenus:2013hs,Larkoski:2014wba, Larkoski:2014uqa, Larkoski:2013paa, Larkoski:2013eya, Dasgupta:2012hg,Dasgupta:2013via,Becher:2015gsa,Chien:2014nsa,Larkoski:2015kga}.

Jet measurements at hadron colliders typically involve identifying jets of size $\cR$ with the use of a jet algorithm, imposing a veto on the out-of-jet transverse momentum $\pTc$ for all radiation\footnote{As  discussed below, to the order  we work this is the same as putting a veto on the third hardest jet.} with (pseudo-)rapidity $y$ in the range $\abs{y} < \yc$ measured with respect to the beam axis. Such measurements are sensitive to hard scales (such as the Mandelstam variables $s,t,u$ in the case of dijet production) in addition to scales induced by the parameters $\cR$, $\yc$, and $\pTc$. When the substructure of jets is probed in the context of a jet measurement,  additional scales such as $Q e$ and $Q e^{\alpha}$ for jet shapes are induced. Thus, there are not only scales associated with the substructure itself but also those associated with the more global context with which the probed jet was produced, and the large set of scales involved can span a wide range of energies. 

Many of the ratios of these scales can be resummed using well known techniques such as SCET in similar ways to those described above for $e^+e^-$. In addition to the ingredients used in $e^+e^-$ collisions, factorization formulae for hadronic collisions involve beam functions $B$ which account for initial-state radiation \cite{Fleming:2006cd,Stewart:2009yx}, and we schematically have
\begin{align}
\label{eq:dsigmapp}
d \sigma^{pp} \sim H \times B \otimes \bar B \otimes J_{n_1}  \otimes \cdots \otimes J_{ n_N} \otimes S_{B \bar{B} n_1 n_2 \cdots}
\,.\end{align}
While RGE of the functions appearing in \eq{dsigmapp} resums a large set of logarithms, others, such as logarithms of $\cR$ \cite{Alioli:2013hba,Kelley:2012zs,Kelley:2012kj} and non-global logarithms (NGLs) \cite{Dasgupta:2001sh,Banfi:2002hw,Appleby:2003ai,Dokshitzer:2003uw}, can present more of a challenge. 
Importantly, resummation of the jet size $\cR$ has recently been explored in the context of subjets in \cite{Dasgupta:2014yra} and in jet rates in the context of $e^+ e^-$ collisions in \cite{Becher:2015hka,Chien:2015cka}, and in addition  there has been progress in understanding NGLs both at fixed-order \cite{Khelifa-Kerfa:2015mma,Hornig:2011tg,Hornig:2011iu, Kelley:2011aa} and more recently a few novel approaches to understanding their all-orders resummation have been proposed \cite{Larkoski:2015zka,Becher:2015hka,Neill:2015nya}.

In this paper we consider the case where the kinematics are such that NGLs are not enhanced and instead focus on resummation of logarithms of ratios of the dynamical scales associated with substructure (such as $Q e/Q$ and $Q e^\alpha/Q$) with fixed $\pTc$, $\yc$, $\cR$, and jet $p_T^J$. To this end, we restrict ourselves to the kinematic region 
\begin{align}
\label{eq:hierarchy}
e&^{-\yc} \ll 1 \nn\\
p_T^J &\sim \sqrt {\hat s} \sim \sqrt{\hat t} \sim \sqrt{\hat u} \nn\\
\pTc \cR^2/p_T^J &\sim e \ll \cR^2 \ll 1
\,.\end{align}
Our approximations are valid to the order we work within about a decade of the value(s) of these parameters for which the NGLs are minimized. In the example we present, we have $e \sim \cO(10^{-3})$ in the peak region of the distribution and $\cR^2 \sim \cO(10^{-1})$, which means the leading NGLs, which are of the form $\as^n \ln^n (\pTc \, \cR^2/p_T^J \, e)$ (and first appear for $n \geq 2$), are not enhanced for $\pTc /p_T^J \sim \cO(10^{-2})$.

One class of event shapes  that has been studied extensively in the literature and is the focus of the present work is that of angularities $\tau_a$, parameterized by a continuous variable $a$ (with $a<2$ for IR safety). The choice $a=0$ corresponds to the classic event shape thrust and $a=1$ corresponds to jet broadening. 
Angularities were originally defined in \cite{Almeida:2008yp,Berger:2003iw} and studied in the context of SCET in \cite{Lee:2006fn,Bauer:2008dt,Hornig:2009vb}. In \Ref{Ellis:2010rwa}, ``jet shapes''\footnote{This is distinct from {\it the jet shape} as defined in \cite{Ellis:1992qq,Ellis:1991vr}  and studied more recently in \Ref{Chien:2014nsa,Chien:2015hda}.} were defined by restricting the angularities to the constituents of a jet as defined by a jet algorithm (as opposed to all particles in the event) and were  resummed to next-to-leading logarithmic (NLL) accuracy. In this work we consider a modified definition of angularities that is designed to be boost invariant about the colliding hadrons' axis, i.e., the beam axis.

We also note  that the definition of the angularities we consider (which differs from that defined for $e^+ e^-$ colliders by a rescaling in the small $\tau_a$ limit) is such that the choice $a=0$ is closely related to the jet mass,
\begin{align}
\tau_{0} = m_J^2/(p_T^J)^2 +  \cO(\tau_0^2)
\,.\end{align}
Jet mass resummation has been studied indirectly by looking at the 1-jettiness global event shape \cite{Stewart:2010tn} for single jet events in \Ref{Jouttenus:2013hs}, by using pQCD methods that neglect color interference effects in \Ref{Dasgupta:2013ihk}, and in the threshold limit in \Refs{Chien:2012ur}{Liu:2014oog}, but to our knowledge has not been studied with the cuts described above, with full NLL' color interference effects\footnote{For an explanation of  which terms are included in our cross section by working to this order, see for example \Ref{Almeida:2014uva}.}, and in a manner that is valid away from the threshold limit. In addition, our results for $a=0$ can be straightforwardly extended to NNLL using the known anomalous dimensions together with the recently deduced two-loop unmeasured jet function anomalous dimension \cite{Chien:2015cka}, which controls the evolution of both unmeasured jet and beam functions. In addition, we apply the refactorization procedure described in  \Ref{Chien:2015cka} which allows the resummation of logarithms of $\cR$ in the region described by \eq{hierarchy}.

While we choose to study angularities as the choice of substructure observable, our basic setup is much more general. Indeed, we obtain many of the results specific to our choice of angularities by using identities that relate the jet functions and the observable-dependent part of our soft function to analogous calculations in  $e^+e^-$ collisions. The part of the soft function that requires an entirely new calculation simply imposes the  experimental $\pTc$ cut on radiation outside of the jets and the beams. This universal part of the soft function, labeled $\vect{S}^{\unmeas}$, encapulates all the interjet cross-talk, and hence contains all perturbative information associated with real emission about the  directions $n_i$ and the color flow. For each jet which has the angularity probed, which here and below we refer to (using the terminology of \Ref{Ellis:2010rwa})  as a ``measured jet," we add a jet function and a soft function contribution that are both angularity dependent but color- and direction-trivial. Thus, other substructure measurements can be straightforwardly incorporated by substituting for their appropriate contributions at this step. If no measurement is performed on a jet (that is, the jet is identified but otherwise unprobed), which we  refer to as an ``unmeasured jet," only an unmeasured jet function (which we also present to $\cO(\as)$) and $\vect{S}^{\unmeas}$ are required. For dijet production, which is the focus of the current work, all four Wilson lines (those of the beams and the two jets) are confined to a plane, and the calculation of $\vect{S}^{\unmeas}$ to $\cO(\as)$ is  tractable. In addition, the effect of different  experimentally used  vetoes, such as putting a $\pTc$ only on the third hardest jet (as opposed to all out-of-jet radiation) will only result in a difference in $\vect{S}^{\unmeas}$ at $\cO(\as^2)$ so our calculations apply there as well.

We also point out that while for unmeasured jets, the jet size $\cR$ must scale with the SCET power counting parameter $\lambda$ and hence the requirement $\cR \ll 1$ is essential, for measured jets this is not strictly needed since $\tau_a \ll 1$ is sufficient to ensure SCET kinematics. 
However, as we will see, both the jet algorithms and measurements simplify significantly  in this limit up to power corrections of the form $\cR^2$ and $\tau_a/\cR^2$, respectively, although we emphasize that the exact results can be obtained numerically using subtractions such as those of \Ref{Bauer:2011hj}. 
Finally, we note that because there is no measurement on any radiation with $\abs{y} < \yc$, our factorization formulae will include ```unmeasured beam functions,'' which to our knowledge have not appeared in the literature.

This paper is organized as follows. In \sec{algs}, we define the classes of jet algorithms and angularity definitions suitable for hadron colliders and relate them to the corresponding $e^+e^-$ algorithms and angularities in the small $\cR$ limit. In \sec{xsection} we outline the $2 \to 2$ kinematic relations needed for dijet production and discuss how both the Born cross section and the fully factorized and resummed SCET cross section are related to the basic building blocks that we then calculate to fixed order in \sec{fixedordercalcs}, namely the hard, jet, soft, and beam functions. We then use these results in \sec{sigmaNLL} to arrive at the NLL' resummed cross section for a generic $2 \to 2$ scattering channel both for when the jets are identified but otherwise left unmeasured (i.e., we are inclusive in the substructure properties) and for when the angularity of either (or both) jets is measured. From our calculations, one can obtain results for the case where  the angularities of both jets $\tau_a^1$ and $\tau_a^2$ are separately measured (and by integrating, the case where $\tau_a^1 + \tau_a^2$ is measured) as well as the cases where only one or neither are measured. For illustrative purposes, in our plots we focus on the case where both $\tau_a^1$ and $\tau_a^2$ are measured and $\tau_a^1 = \tau_a^2$.
Furthermore, we present explicit results for the simple channel $qq' \to qq'$ with different values of $\cR$ and $p_T^{\rm cut}$ and for several choices of the angularity parameter $a$, and demonstrate the reduction in scale uncertainty resulting from the refactorization techniques of \cite{Chien:2015cka}. We conclude in \sec{conclusion}.

\section{Jet Algorithms and Shapes at Hadron Colliders}
\label{sec:algs}

The main difference between jet cross section measurements at $e^+e^-$ colliders and hadron colliders is that the latter prefer observables that are invariant under boosts along the beam direction.
The $k_T$-type algorithms used at the LHC (described in more detail in, for example, Ref.~\cite{Salam:2009jx}) merge particles successively using a pairwise metric
\begin{align}
\rho_{ij} = {\rm min}\{(p_T^i)^{2p}, (p_T^j)^{2p} \} \frac{\Delta \cR_{ij}^2}{\cR^2}
\,,\end{align}
where $p = +1, 0$, and $-1$ for the $k_T$, C/A, and anti-$k_T$ algorithms, respectively, $p_T^i$ is the transverse momentum (with respect to the beam) of particle $i$, $\cR$ is a parameter characterizing the jet size, and 
\begin{align}
\label{eq:cR}
\Delta \cR_{ij} \equiv \sqrt{(\Delta y_{ij})^2 + (\Delta \phi_{ij})^2}
\,,\end{align}
where $\Delta y_{ij}$ and $\Delta \phi_{ij}$ are the pseudo-rapidity and azimuthal angle differences of the particles measured with respect to the beam axis. Since pseudo-rapidities simply shift under boosts and azimuthal angles are invariant, $\Delta \cR_{ij}$ is invariant under boosts along the beam direction.
This pairwise metric is  compared to the single particle metric of each particle, defined as
\begin{align}
\rho_i = (p_T^i)^{2p}
\,.\end{align}
Two particles are merged if their pairwise metric is the smallest for the $(ij)$ pair over all particle pairs and is less than both of the single particle metrics, i.e., $\rho_{ij} < {\rm min}\{ \rho_i, \rho_j \}$. This latter constraint amounts to 
\begin{align}
 \Delta \cR_{ij} < \cR
\,.\end{align}
In the following, we will work under the assumption that all particles in the jet are close to a jet axis at polar angle $\theta_J$ with respect to the beam axis such
that $\Delta\cR_{ij}$ can be expanded as
\begin{align}
\label{eq:smallcR}
 \Delta \cR_{ij} &= \frac{1}{\sin \theta_J} \sqrt{(\Delta \theta_{ij})^2 + \sin^2\theta_J (\Delta \phi_{ij})^2} + \cO((\Delta \theta_{ij})^2, (\Delta \phi_{ij})^2)\nn\\
 &= \frac{\theta_{ij}}{\sin \theta_J} + \cO(\theta_{ij}^2)
\,,\end{align}
where in the first equality $\Delta \theta_{ij}$ and $\Delta \phi_{ij}$ are the angle differences in a spherical coordinate system  with $\hat z$ in the  beam axis direction, and $\theta_{ij}$ in the second equality is simply the angle between particles $i$ and $j$. 
This implies we can impose an $e^+e^-$-type polar angle restriction that particles are within a jet of size $R$ and rescale the results by 
\begin{align}
\label{eq:rescale} 
R \to \cR \sin \theta_J = \frac{\cR}{\cosh y_J}
\,,\end{align}
where $y_J$ is the jet pseudo-rapidity, up to $\cO(\cR^2)$ corrections.
This allows us to recycle many of the results of \Ref{Ellis:2010rwa}. The difference between our results and those obtained from the exact expression \eq{cR} can be obtained numerically, e.g., with the methods of \Ref{Bauer:2011hj}, although the details are beyond the scope of the present work.

It is helpful to re-write the angularity definition used in \Ref{Ellis:2010rwa} in the context of $e^+e^-$ collisions in terms of ingredients that are boost invariant, such as $p_T$ and the right-hand side of \eq{smallcR}. To do so, first recall the definition used in terms of the pseudo-rapidities $y_{iJ}$ and transverse momenta $p_{\perp}^{iJ}$ of particles {\it with respect to the jet axis},
\begin{align}
\tau_a^{e^+e^-} &= \frac{1}{2 E_J} \sum_{i \in {\rm jet}} \vert p_{\perp}^{iJ} \vert  e^{-(1-a) \abs{y_{iJ}} }
\,.\end{align}
In the small angle approximation, we can write this as
\begin{align}
\tau_a^{e^+e^-} &= (2 E_J)^{-(2-a)} (\pTJ)^{1-a} \sum_{i \in {\rm jet}}  \vert p_{T}^{i} \vert  \bigg( \frac{\theta_{iJ}}{\sin\theta_{J}} \bigg)^{2-a} \big(1+ \cO(\theta_{iJ}^{2})\big)
\,.\end{align}
From the discussion above, all terms in the sum over particles are boost invariant. The one term that is not boost invariant is just the overall factor of $(2E_J)^{2-a}$. Therefore, we can arrive at a boost invariant version of $\tau_a$ suitable for hadron colliders with a simple rescaling by a dimensionless factor,
\begin{align}
\label{eq:taupp}
\tau_a \equiv \tau_a^{pp} &\equiv \frac{1}{\pTJ} \sum_{i \in {\rm jet}}   \vert p_{T}^{i} \vert (\Delta \cR_{iJ})^{2-a} \nn\\
&= \bigg(\frac{2 E_J}{\pTJ}\bigg)^{\!2-a} \tau_a^{e^+e^-} + \cO(\tau_a^2)
\,.\end{align}
We emphasize again that the quantities on the right-hand side of the first line of \eq{taupp} are manifestly invariant under boosts along the beam axis, and that the second line allows us to recycle many of the results of \Ref{Ellis:2010rwa}. 

The one main difference between measurements done at $e^+e^-$ colliders and  hadron colliders that requires a novel calculation is the out-of-jet energy veto. In $e^+e^-$ colliders, this is typically  a cut on energy, whereas in hadron colliders it is typically a veto on transverse momentum: $p_T = E \sin\theta < p_T^{\rm cut}$. This will require an entirely new soft function, which we present below.

\section{Factorized Dijet Cross Section}
\label{sec:xsection}

For dijet production at tree-level, momentum conservation implies that there are just three non-trivial variables to describe the final state at tree level, which we can take to be the jet (pseudo-) rapidities $y^{1,2}$ and the jet $\pTJ = \vert  \vect{p}_T^1\vert = \vert  \vect{p}_T^2 \vert$. The momentum fractions of the incoming partons are related to these variables via
\begin{align}
\label{eq:x12}
x_{1,2} &= \frac{2 \pTJ }{\Ecm} \cosh \frac{\Delta y}{2} e^{\pm Y}
\,,\end{align}
where  $\Delta y = y_1 - y_2$ is the rapidity difference of the two jets and $Y = (y_1+y_2)/2$. The (partonic) Mandelstam variables can be written as
\begin{align}
\label{eq:mandelstam}
s & = 4 \pTJ^2 \cosh^2 \frac{\Delta y}{2} \nn\\
t &= - 2 \pTJ^2 e^{\Delta y/2} \cosh  \frac{\Delta y}{2} \nn\\ 
u &= - 2 \pTJ^2 e^{-\Delta y/2} \cosh  \frac{\Delta y}{2} = -s-t
\,.\end{align}

The tree-level matrix element squared can be written as
\begin{align}
\vert \mathcal{M}_{\rm tree} \vert^2 =   \Tr \{ \vect H_0 \vect S_0\} \, ,
\end{align}
where $\vect H_0$ and $\vect S_0$ are the tree-level hard and soft functions, respectively, so the Born cross section takes form 
\begin{align}
\label{eq:born}
\frac{d \sigma_{\born}}{d y_1 d y_2 d \pTJ} = \Norm f_1(x_1,\mu)f_2(x_2, \mu)  \Tr \{ \vect H_0 \vect S_0\}
\end{align}
where $\norm$ is the normalization associated with averaging over initial particle quantum numbers (e.g., $\norm = 4 N_c^2$ for quark scattering) and $ f_i(x_i,\mu)$ is a PDF for parton $i$ with momentum fraction $x_i$.

The effect of radiative corrections to \eq{born} is described in the soft and collinear limits by  higher-order hard, soft, beam, and jet functions. We consider the cases when both jets are unmeasured and when both jets are measured. 
When both jets are unmeasured the all-orders cross section takes the form
\begin{align}
\label{eq:sigmafactunmeas}
d \sigma &\equiv \frac{d \sigma}{d y_1 d y_2 d \pTJ } \\
&=\Norm \Beam (x_1, \mu) \Beambar(x_2, \mu) 
 \Tr \{ \vect H(\mu) \vect S^{\unmeas} (\mu)\}  J_1(\mu) J_2(\mu) \nn\\
 & \qquad + \cO(\as \cR^2, \as e^{-2\yc} ) 
\,,\end{align}
where the $J_i(\mu)$ are unmeasured jet functions and $S^{\unmeas}$ is the unmeasured soft function. 
When both jets are measured, the cross section takes the form
\begin{align}
\label{eq:sigmafactmeas}
d \sigma(\tau_a^1, \tau_a^2) &\equiv \frac{d \sigma}{d y_1 d y_2 d \pTJ d \tau_a^1 d \tau_a^2} \\
&= \Norm \Beam (x_1, \mu) \Beambar(x_2, \mu) 
\Tr \{ \vect H(\mu) \vect S(\tau_a^1, \tau_a^2, \mu)\} \otimes [J_1(\tau_a^1, \mu) J_2(\tau_a^2, \mu)] 
\nn\\ & \qquad 
+ \cO(\as \tau_a^{i}/\cR^2, \as e^{-2\yc}) \, , \nn
\end{align}
where $\otimes$ represents the two convolutions over the $\tau_a^{1,2}$. The case of a single measured jet, with the other jet unmeasured, is the obvious generalization of \eqs{sigmafactunmeas}{sigmafactmeas}. 
The power corrections to \eqs{sigmafactunmeas}{sigmafactmeas} can be included via matching to fixed order QCD. Resummation of logs of $\tau_a$ is achieved by RG evolution of each factorized component from its canonical scale (cf. \tab{anomalous-coeff}) to the common scale $\mu$. Both the hard and soft function  are in general  matrices (which here and below we will refer to with bold face) which are hermitian and of rank $R$ equal to the number of linearly independent color operators associated with the hard process (e.g., $R=2$ for $qq \to qq$, 3 for $qq \to gg$, and 8 for $gg \to gg$). These operators mix under RG evolution which is accounted for with matrix RG equations. The fixed order calculation of the components in \eqs{sigmafactunmeas}{sigmafactmeas} and their RG evolution is the subject of the next sections. 

\section{Fixed-Order $\cO(\as)$ Calculation of Factorized Components}
\label{sec:fixedordercalcs}

\subsection{Jet Functions}
\label{ssec:jet}

In \Ref{Ellis:2010rwa}, there are both ``measured'' and ``unmeasured'' jet functions, corresponding to jets whose angularity was measured as opposed to those that were identified but otherwise unprobed. The latter can be obtained using the hadron collider algorithms with the rescaling in \eq{rescale}. We obtain
\begin{align}
\label{eq:bareunmeasjet}
J_i &= 1 + \frac{\as}{2 \pi}\bigg[ \bigg(  \frac{\cas{i}}{\epsilon^2} + \frac{\gamma_i}{\epsilon} \bigg) \bigg(\frac{\mu}{\pTJ \cR}\bigg)^{\!2\epsilon}  + d_J^{i, {\rm alg}}\bigg] 
\end{align}
where $i=q,g$ for quark and gluon jets (and $\cas{i}$ is the Casimir invariant, $C_q = \CF$ and $C_g = \CA$), respectively, and 
\begin{align}
\label{eq:gammaqg}
\gamma_q = \frac{3 \CF}{2} \ , \quad \gamma_g = \frac{\beta_0}{2}\,.
\end{align}
(with $\beta_0$ given  in \eq{beta-pieces})
and the finite corrections $d_J^{i, {\rm alg}}$ are given in Eqs. A.19 and A.30 of \cite{Ellis:2010rwa},
\begin{align}
d_J^{i, {\rm cone}} =  2\gamma_i  \log 2 -\cas{i} \frac{5\pi^2}{12}+&
\begin{cases}
\CF\frac{7}{2} &\mbox{if } i=q\\
\CA\frac{137}{36}-\TR\NF\frac{23}{18} &\mbox{if } i=g
\end{cases} \\
d_J^{i, k_T} = - \cas{i} \frac{3\pi^2}{4}+&
\begin{cases}
\CF\frac{13}{2} &\mbox{if } i=q\\
\CA\frac{67}{9}-\TR\NF\frac{23}{9} &\mbox{if } i=g
\end{cases} 
\end{align} 
where $d_J^{i, k_T}$ is the same constant for all $k_T$-type algorithms ($k_T$, anti-$k_T$, and C/A).

For measured jet functions, we need to apply the rescaling \eq{taupp}. The identity
\begin{align}
A^{-1} \delta \big(A^{-1} \tau- \hat \tau \big) = \delta\big(\tau - A \hat \tau\big) 
\,,\end{align}
implies that this rescaling can be accomplished to all orders via the transformation
\begin{align}
J_i(\tau_a) &=  \bigg(\frac{\pTJ}{2 E_J}\bigg)^{\!2-a} J^{e^+e^-}_i \bigg(\bigg(\frac{\pTJ}{2 E_J}\bigg)^{\!2-a}  \tau_a\bigg)
\,,\end{align}
where $J^{e^+e^-}_i(\tau_a)$ is the jet function of \cite{Ellis:2010rwa}. This gives
\begin{align}
J_i(\tau_a) &= J^{e^+e^-}_i(\tau_a)\big\vert_{2 E_J \to \pTJ}
\,,\end{align}
i.e., it is simply obtained from $J^{e^+e^-}_i(\tau_a)$ by making the replacement $2 E_J \to \pTJ$. 
These can be obtained for the quark case from \Ref{Hornig:2009vb} and for the gluon case 
by performing the integral in Eq. (4.22) of \Ref{Ellis:2010rwa}  after  setting $\Theta_{\rm alg}(x) \to 0$ which is valid to $ \cO(\tau_a/\cR^2)$.
We record the results here as
\begin{align}
\label{eq:jetFO}
J_i(\tau_a) &= \delta(\tau_a) - \frac{\as}{2 \pi} \bigg[ \bigg(\frac{\mu}{\pTJ}\bigg)^{\!2\epsilon} \bigg(\frac{1}{\tau_a}\bigg)^{\!1+\frac{2\epsilon}{2-a}}\bigg(\frac{1}{\epsilon}\frac{2\cas{i}}{1-a} + \frac{\gamma_i}{1-a/2}\bigg)- \delta(\tau_a) f_i(a) \bigg] 
\,,\end{align}
where
\begin{align}
\label{eq:fia}
f_q(a) &= \frac{2 \CF}{1-a/2}\bigg[ \frac{7-13a/2}{4} - \frac{\pi^2}{12}\frac{3-5a+9a^2/4}{1-a} \\
& \qquad \qquad \quad - \int_0^1\!dx\, \frac{1-x+x^2/2}{x}\log[x^{1-a} + (1-x)^{1-a}] \bigg]\nn\\
f_g(a) &=  \frac{1}{1-a/2}\bigg[ \CA \bigg( (1-a) \bigg(\frac{67}{18}-\frac{\pi^2}{3}\bigg)+\frac{\pi^2}{6} \frac{(1-a/2)^2}{1-a}\nn\\
& \qquad\qquad \quad - \int_0^1\!dx\,   \frac{(1-x(1-x))^2}{x(1-x)}\log[x^{1-a} + (1-x)^{1-a}]  \bigg) \nn\\
& \quad - \TR\NF \bigg( \frac{20-23 a}{18}  - \int_0^1\!dx\,  \big(2x(1-x)-1 \big)\log[x^{1-a} + (1-x)^{1-a}] \bigg) \bigg] \nn
\,.\end{align}
Finally, we note that the integral over $\tau_a$ of the measured jet function is not simply related to the unmeasured jet function and refer the reader to Ref.~\cite{Chien:2015cka} for a detailed explanation.

\subsection{Unmeasured Beam Functions}
\label{ssec:beam}

While the unmeasured beam function has not to our knowledge appeared in the literature, it is directly related to the unmeasured fragmenting jet function of \cite{Procura:2011aq}. The unmeasured fragmenting jet function for a jet of energy $E$ and ($e^+e^-$) cone radius $R$ can be written as
\begin{align}
\label{eq:cG}
\cG(E,R, z, \mu) =\sum_i \int\!\frac{dz'}{z'} \cJ_{ij}(E,R, z', \mu) D_j^h(z/z', \mu) + \cO(\Lambda_{\rm QCD}^2/E^2)
\,,\end{align}
where $D_i^h(x, \mu)$ is a fragmentation function for parton $i$ in hadron $h$ and the $\cJ_{ij}$ are matching coefficients which are given in Eq.~(5) of \Ref{Procura:2011aq}. The dependence on $E$ and $R$ in $\cJ_{ij}$ (at least to $\cO(\as)$) is such that we can write
\begin{align}
\cJ_{ij}(E,R, z', \mu)  \equiv \cJ_{ij}(2 E \tan\frac{R}{2},z', \mu) 
\,,\end{align}
i.e., $E$ and $R$ always appear in the combination $E \tan\frac{R}{2}$. 
Using the crossing relations of Sec. IIIC of \Ref{Ritzmann:2014mka}, it can be shown that an unmeasured beam function in a collider with center-of-mass energy $\Ecm$ and a rapidity cut of $\yc$ can be written as  
\begin{align}
\label{eq:BB}
B_i(x_i, \mu) &\equiv B_i(\Ecm, \yc, x_i, \mu) \nn\\
&= \sum_j \int\!\frac{dz}{z} \cJ_{ij}(x_i \Ecm e^{-\yc}, z, \mu) f_j(x_i/z, \mu) + \cO(\Lambda_{\rm QCD}^2/E^2)
\,\end{align}
where  $\cJ_{ij}$ are {\it the same} matching coefficients as in \eq{cG}, at least to $\cO(\as)$,\footnote{It is argued in \cite{Procura:2009vm} that measured beam and jet functions have the same anomalous dimension to all orders (at least for the measured case), but since the PDFs and fragmentation functions differ perturbatively at $\cO(\as^2)$ \cite{Ellis:1991qj} the matching coefficients must differ for the beam and jet functions starting at this order.} and we used the correspondence between an $e^+e^-$ jet and a beam with label momentum $x_i\Ecm$ and rapidity cut $\yc$
\begin{align}
E \tan\frac{R}{2} \to x_i\Ecm e^{-\yc} 
\,,\end{align}
which is valid up to $\cO(e^{-2\yc} )$ corrections. For the dijet cross section we consider, the $x_i$ are fixed via \eq{x12}.


\subsection{Soft Function}
\label{ssec:soft}

In general, we can write the bare soft function at $\cO(\as)$ for dijet production when both jets have $\tau_a$ measured as
\begin{align}
\label{eq:meas+unmeas}
\vect S(\tau_a^1, \tau_a^2) &= \vect S^{\rm unmeas}\delta(\tau_a^1) \delta(\tau_a^2) 
+ [\vect S_0 S^{\rm meas}(\tau_a^1) \delta(\tau_a^2) + (1 \leftrightarrow2)] + \cO(\as^2)
\,,\end{align}
where  $ \vect S^{\rm unmeas} = \vect S_0 + \cO(\as)$ is the part of the soft function that is always present (both when the jets are measured and unmeasured). The bare soft function is $\mu$ independent, and we will distinguish the corresponding renormalized function with an explicit argument $\mu$.
In the cases that neither of the jets or only one jet is measured, the corresponding $S^{\rm meas}$ pieces on the right-hand are simply not included, while $\vect{S}^{\unmeas}$ is always included. For more jets, the result can be extended straightforwardly, although our explicit results only apply to planar jet configurations (as is  necessarily the case for dijet production).

\subsubsection{Calculation of the One-Loop Ingredients}
\label{sec:oneloopingredients}

The part of the soft function corresponding to the measurement of $\tau_a^i$ on jet $i$, $S^{\rm meas}(\tau_a^i)$, is obtained from summing over the interference of jet $i$ with all other jets and the beams.  Contributions from radiation arising from the interference of jets/beams $j$ and $k$ with $j,k \neq i$ give power corrections in $\cR$. The calculation of $S^{\rm meas}(\tau_a^i)$ can be obtained from the results for $S_{ij}^{\rm meas}(\tau_a^i)$ given in Eq. (5.18) of \Ref{Ellis:2010rwa} through the rescaling in \eq{taupp}. We find
\begin{align}
\label{eq:smeas}
S^{\rm meas}(\tau_a^i) &= 2 \sum_{\pairs} \bigg(\frac{\pTJ}{2 E_J}\bigg)^{\!2-a}  S^{\rm meas}_{ij} \bigg(\bigg(\frac{\pTJ}{2 E_J}\bigg)^{\!2-a} \tau_a^i\bigg)\nn\\
&=  \frac{1}{\epsilon} \frac{\as \cas{i}}{\pi} \frac{e^{\gamma_E \epsilon}}{\Gamma(1-\epsilon)}\frac{1}{1-a}\bigg(\frac{1}{\tau_a^i}\bigg)^{1+2\epsilon} \bigg(\frac{\mu}{\pTJ} \bigg)^{\!2\epsilon}\cR^{2\epsilon(1-a)}
\,,\end{align}
which clearly has the desired boost-invariant properties.

The additional part of the soft function we require, $\vect S^{\rm unmeas}$, can be written as a sum of contributions in the same manner as \Ref{Ellis:2010rwa},
\begin{align}
\label{eq:Sunmeas-defn}
\vect S^{\rm unmeas}&= \vect S_0 + \bigg[ \vect S_0\sum_{\pairs} \Tij \Big( S_{ij}^{\rm incl} + \sum_{k=1}^N S_{ij}^k\Big) + \hc \bigg]
\,,\end{align}
where $\hc$ denotes the hermitian conjugate. Here, we use the color space formalism as described in Refs.~\cite{Catani:1996jh,Catani:1996vz}. The $4!/(2!)^2 = 6$ matrices $\Tij$ are of rank $R$, the same as that of $\vect S_0$, and account for the mixing of color operators in a given basis into each other at $\cO(\as)$.
The difference from \Ref{Ellis:2010rwa} is that now each contribution involves a $p_T$ veto instead of an energy veto as well as a different jet algorithm. 
In particular, defining
\begin{align}
\label{eq:thetas}
 \Theta_{p_T} &\equiv  \Theta(k^0 \sin\theta_{kB}<p_T^{\rm cut}) \nn\\
  \Theta_\cR^k  &\equiv \Theta( \cR_{kJ} < \cR)
\,,\end{align}
we now have 
\begin{align}
\label{eq:Sincl-integral}
S_{ij}^{\incl} \equiv  \frac{1}{\epsilon} \frac{\as}{2\pi}\bigg(\frac{\mu}{\pTc}\bigg)^{\! 2\epsilon} \, \cI^{\incl}_{ij}= - g^2 \mu^{2 \epsilon}   \!\int\!\frac{d^dk}{(2 \pi)^{d-1}} \frac{n_i \cdot n_j}{(n_i \cdot k) (n_j \cdot k)} \delta(k^2)\Theta(k^0) \Theta_{p_T}
\,,\end{align}
and
\be
\label{eq:Sijk}
S_{ij}^k \equiv  \frac{1}{\epsilon} \frac{\as}{2\pi}\bigg(\frac{\mu}{\pTc}\bigg)^{\! 2\epsilon} \, \cI_{ij}^k = g^2\mu^{2\epsilon} \!\int\!\frac{d^dk}{(2\pi)^{d-1}}\frac{n_i\cdot n_j}{(n_i\cdot k)(n_j\cdot k)}\delta(k^2)\Theta(k^0) \,  \Theta_{p_T} \Theta_\cR^k
\,,\ee
where $i$, $j$, and $k$ can each be  either of the beams or one of the jets (with $i \neq j$).

We first perform the energy and trivial parts of the angular integration of \eq{Sincl-integral} for generic $i,j$ (either jet or beam). To do this, we align the $1$-direction (or ``$\hat z$'') with direction $\vec n_i$ and put the $\vec n_j$ vector in the 12-plane, and the beam direction $\vec n_B$ in the 123-spatial part of $d$-dimensional space. Using the shorthands $c_{ij} \equiv 1-n_i \cdot n_j$, $s_{ij} \equiv (1-c_{ij}^2)^{1/2}$, $c_i \equiv \cos\theta_i$, and $s_i \equiv \sin\theta_i$,  the dot products of the gluon's 3-momentum, $\vec k$, with these unit vectors take the form
\begin{align}
\vec n_i \cdot \vec k &= c_1 \nn\\
\vec n_j \cdot \vec k &= c_{ij} c_1 + s_{ij} s_1 c_2 \nn\\
\vec n_B \cdot \vec k &=  n_{B1} c_1 + n_{B2} s_1 c_2 + n_{B3} s_1 s_2 c_3 \, ,
\end{align}
for the $i$, $j$, and beam directions, respectively. In this frame, $\cI_{ij}^{\rm incl}$ takes the form (in $\overline {\rm MS}$)
\begin{align}
\label{eq:Iij-integral}
\cI^{\incl}_{ij} &= \frac{(1-c_{ij})e^{\gamma_E \epsilon}}{2 \sqrt \pi \Gamma(1/2-\epsilon)} \int_0^\pi\!d\theta_1\, \sin^{1-2\epsilon}\theta_1 \int_0^\pi \!d\theta_2\, \sin^{-2\epsilon}\theta_2 \frac{1}{1-c_1} \frac{1}{1-c_{ij} c_1 - s_{ij} s_1 c_2} \nn\\
& \quad \times \!\!\bigg[ \frac{\Gamma(1/2-\epsilon)}{\sqrt \pi \Gamma(-\epsilon)}\int_0^\pi\!d\theta_3\, \sin^{-1-2\epsilon}\theta_3 \big(1-(n_{B1} c_1 + n_{B3} s_1 c_2 + n_{B3} s_1 s_2 c_3)^2\big)^\epsilon \bigg] 
\,.\end{align}
The quantity in parenthesis to the $\epsilon^1$ power in the second line is the square of the sine of the gluon-beam angle and comes from doing the $k^0$ (energy) integral over the $p_T$ veto, $\Theta_{p_T}$. For planar events (such as dijet events at hadron colliders), $n_{B3}=0$ (since the beam is in the $ij$-plane for all $i,j$) and the integration over $\theta_3$ can be easily performed. The entire second line (the quantity in brackets) then becomes simply
\begin{align}
\label{eq:planar}
\bigg[ \cdots \bigg] \xrightarrow{{\rm planar}} \big(1-(n_{B1} c_1 + n_{B2} s_1 c_2)^2\big)^\epsilon
\,,\end{align}
with $n_{B2}^2 = 1-n_{B1}^2$.
We also note that when $i$ is equal to the beam direction (so $n_{B1}=1$ and $n_{B2}=0$), this quantity reduces to 
\begin{align}
\bigg[ \cdots \bigg] \xrightarrow{n_i = n_B} \sin^{2\epsilon}\theta_1
\,.\end{align}
In this case, the $\epsilon$ dependence in the overall power of $\sin\theta_1$ cancels and we are left with a divergence unregulated by dimensional regularization. This is the well-known rapidity divergence  that is present for a $p_T$ veto. This can be treated within the context of ${\rm SCET}_{\rm II}$ as was done for example in \Ref{Chiu:2011qc}. Here, we will opt instead  to veto on radiation only below a rapidity cut $\yc$ which is consistent with what is done at the LHC since radiation going down the beam pipes is not measured.
We compute the soft function components $\cI_{ij}^i$ and $\cI_{ij}^{\rm incl}$ for the case $i$ and $j$ can each either be beams or jets in  \app{app:soft} and record the results in Table \ref{tab:softresults}. For the case that either $i$ or $j$ is a beam, we only compute the full out-of-beam contribution, e.g.  $\cI_{JB}^{\rm incl} +\cI_{JB}^B$ (or  $\cI_{B\Bbar}^{\rm incl} +\cI_{B\Bbar}^B + \cI_{B\Bbar}^{\Bbar}$ for the case both $i$ and $j$ are beams) to avoid having to regulate the rapidity divergences in individual components.

\TABLE[b]{
\begin{tabular}{||c | c @{}c ||}
\hline\hline
\tabrule
contribution & result & \\
\hline\hline
\tabrule
$ \qquad \cI_{B\Bbar}^{\incl} + \cI_{B\Bbar}^B + \cI_{B\Bbar}^{\Bbar} \qquad$ & $ 2 \yc $ & \\
\tabrule
$\cI_{B\Bbar}^1 + \cI_{B\Bbar}^2$ & $\cO(\cR^2)$ & \\
\hline
\tabrule
$\cI_{B J}^{\incl} + \cI_{BJ}^B + \cI_{B J}^{\Bbar}$ & $ -\frac{1}{2\epsilon}  + \yc -y_J  + \epsilon \frac{\pi^2}{24}$ & \\
\tabrule
$\cI_{B J}^{J}$ & $\frac{1}{2\epsilon} \cR^{-2\epsilon} \big(1- \epsilon^2 \frac{\pi^2}{12}\big)$  &\\
\tabrule
$\cI_{B J}^{k \neq J,B}$ & $\cO(e^{-\yc}, \cR^2)$ &  \\
\hline
\tabrule
$\cI_{12}^{\incl}$ &  $ (2 \cosh \frac{\Delta y}{2})^{-2\epsilon}(-\frac{1}{\epsilon}  +\frac{\epsilon}{2}(\Delta y)^2  + \epsilon \frac{\pi^2}{12})$  &\\
\tabrule
$\cI_{12}^{1}+\cI_{12}^{2}$ &  $ \frac{1}{\epsilon}  \cR^{-2\epsilon}  \big(1- \epsilon^2 \frac{\pi^2}{12}\big)$  &\\
\tabrule
$\cI_{12}^{B,\Bbar}$ &  $\cO(e^{-\yc})$  &\\
\hline\hline
\end{tabular}
\caption{A summary of results for the ``unmeasured'' part of the soft function, $S^{\rm unmeas}$, up to $\cO(e^{-\yc}, \cR^2)$. Here, the subscript $J$ refers  to the two jets, $J=1,2$, and $B$ and $\Bbar$ refer to the two beams, and $\Delta y = y_1-y_2$. Each component is explicitly boost invariant about the beam direction (with $2\yc$ in the $B$-$\Bbar$ interference terms in general given by the rapidity difference of the forward and backward beam cuts). }
\label{tab:softresults}
}

For several of the components, we use the fact that the result is boost invariant along the beam direction to boost to the frame where the jets are back-to-back. The relation between the back-to-back frame beam-jet angle $\theta_J$ and the jet rapidities in the lab frame is
\begin{align}
\label{eq:b2bframeangle}
\cos \theta_J = \tanh \frac{\Delta y}{2}
\,,\end{align}
where $\Delta y = y_1 - y_2$ is the rapidity difference of the two jets. This also means that when putting a polar angle restriction on the emitted gluon in the back-to-back frame, one has to apply the correspondence \eq{b2bframeangle}  in using \eq{rescale}, which amounts to the replacement 
\begin{align}
\label{eq:b2bframerescale}
\tan \frac{R}{2} \to \frac{\cR}{2\cosh \Delta y/2}
\,,\end{align}
where dependence on the left-hand side arises from enforcing a restriction on the polar angle of the gluon about a jet ($\theta<R$) in the back-to-back frame.

Using the  color algebra identity $\sum_i \vect T_i =0$ and the kinematic relations 
\begin{align}
 \log \frac{n_J \cdot n_B}{2}&= - y_J  - \log(2 \cosh y_J)\nn\\
\log \frac{n_J \cdot \bar n_B}{2}&=  y_J  - \log(2 \cosh y_J)
\,,\end{align}
for jets $J=1,2$, and
\begin{align}
\ln\frac{n_1 \cdot n_2}{2} &= \ln\frac{(2 \cosh \Delta y/2)^2}{(2\cosh y_1)(2\cosh y_2)}
\,,\end{align}
we find
\begin{align}
\label{eq:Sunmeas}
\vect S^{\unmeas} &=\vect S_0 + \frac{\as}{\pi} \bigg\{ \vect S_0 \bigg[ \bigg( \frac{1}{2\epsilon} +  \log \frac{\mu}{\pTc}\bigg) \Big( \vect S^{\rm div}+  \sum_{i=1,2} \cas{i} \ln \cR\Big)-\frac{1}{2} \sum_{i=1,2} \cas{i}\log^2 \cR \nn\\
& \qquad  \qquad \qquad  \qquad -  \vect T_1 \mcdot \vect T_2 \log\big(1+ e^{\Delta y}\big)\log\big(1+ e^{-\Delta y}\big)\bigg] + \hc \bigg\} + \cO(\as^2)
\,.\end{align}
In this equation,
\begin{align}
\label{eq:Sdiv}
\vect S^{\rm div} &=\sum_{\pairs} \Tij \log \frac{n_i \cdot n_j}{2} - \yc \big( \cas{B}+ \cas{\Bbar})
-  \sum_{i=1,2}\cas{i} \log (2 \cosh y_i)  \nn\\
&= \gammaunmeas(m_i)  -\Ms(m_i)
\,,\end{align}
where in the second line we wrote the result in terms two functions defined by 
\begin{align}
\label{eq:SdivD}
\gammaunmeas(m_i) &= \sum_{i=B,\Bbar} \cas{i}  \ln\frac{x_i \Ecm e^{-\yc}}{m_i} + \sum_{i=1,2}  \cas{i} \ln\frac{\pTJ }{m_i} \nn\\
\Ms(m_i) &\equiv - \sum_{\pairs}\Tij \ln\frac{s_{ij}}{m_i m_j}
\,,\end{align}
where $s_{ij} \equiv 2p_i\cdot p_j >0$ (and where $p_i = x_i \Ecm$ for the beams $i = B, \bar{B}$). Note that for later convenience we have defined these functions so that each separately depends on a set of parameters $m_i$. 
The dependence on $m_i$ cancels in the sum in the second line of Eq.~(\ref{eq:Sdiv}).

\subsubsection{Refactorization}
\label{sec:refact}

We note here that one can also construct the ingredients needed for the refactorized cross section as was done in \Ref{Chien:2015cka} for the resummation of (global) logs of $\cR$ from the ingredients in \tab{softresults}. In particular, the conclusions of  \Ref{Chien:2015cka} suggest that $\vect S^{\unmeas}$ should be factorized as 
\begin{align}
\label{eq:Sunmeasrefact}
\vect S^{\unmeas} &= \frac{1}{2} \vect{S}_0 \int_0^{\pTc} dE \, \Big[\vect{s}_{s}(E) \otimes {s}_{sc}^1(E \cR)  \otimes {s}_{sc}^2(E \cR)\Big] + \hc  \nn\\
&= \vect{S}_0 + \frac{\as}{4\pi}\frac{1}{2} \Big[ \vect{S}_0  \Big( \vect{S}_{s}^{(1)}(\pTc) + \sum_{k=1,2}S_{sc}^{k(1)} (\pTc \cR)\Big) + \hc \Big] + \cO(\as^2)
\,,\end{align}
where $\otimes$ is a convolution over the variable $E$ and the functions $\vect{S}_{s}$ and $S_{sc}^k$ are the global soft (with radiation anywhere except for the beams) and soft-collinear  (with radiation within jet $k$) functions, respectively, and where
\begin{align}
\vect{s}_s (\pTc) &\equiv \frac{d}{d \pTc}  \vect{S}_s(\pTc)  \nn\\
{s}_{sc}^k (\pTc \cR) &\equiv \frac{d}{d \pTc} S_{sc}^k(\pTc \cR) 
\end{align}
with both functions $f = \vect{S}_s, S_{sc}^k$ normalized as $f(x) = \theta(x) + \sum_{i=1} (\frac{\as}{4\pi})^n f^{(n)}(x)$. Note that all of the non-trivial color mixing occurs in $\vect{S}_s$. This is due to the fact that the soft-collinear modes of  \Refs{Becher:2015hka}{Chien:2015cka} are confined to a single jet and is expected to hold to all orders.

In terms of the ingredients in \tab{softresults}, we have 
\begin{align}
\label{eq:Sss1}
\vect{S}_{s}^{(1)}(\pTc) &= \frac{4}{\epsilon} \bigg(\frac{\mu}{\pTc} \bigg)^{\! 2\epsilon} \sum_{i<j} \vect{T}_i \cdot \vect{T}_j \bigg[ \cI_{ij}^{\incl} + (\delta_{iB} + \delta_{i \Bbar})(\delta_{j J_1} + \delta_{j J_2}) \cI_{ij}^i + \delta_{iB} \delta_{i \Bbar} (\cI_{ij}^i+\cI_{ij}^j)\bigg] \nn\\
&=  \frac{4}{\epsilon} \bigg(\frac{\mu}{\pTc} \bigg)^{\! 2\epsilon}  \bigg[ \sum_{i=1,2}\frac{C_i}{2\epsilon} \Big(1 - \epsilon^2 \frac{\pi^2}{12} \Big) + \vect S^{\rm div} - 2\epsilon \, \vect T_1 \mcdot \vect T_2 \log\big(1+ e^{\Delta y}\big)\log\big(1+ e^{-\Delta y}\big)\bigg]
\end{align}
and 
\begin{align}
\label{eq:Ssc1}
S_{sc}^{k(1)}(\pTc \cR) &= \frac{4}{\epsilon} \bigg(\frac{\mu}{\pTc} \bigg)^{\! 2\epsilon} \sum_{i<j} \vect{T}_i \cdot \vect{T}_j \Big[ \delta_{ik}\cI_{ij}^i \Big] 
= 
\frac{4}{\epsilon} \bigg(\frac{\mu}{\pTc \cR} \bigg)^{\! 2\epsilon} \Big[- \frac{C_k}{2\epsilon} \Big(1 - \epsilon^2 \frac{\pi^2}{12} \Big) \Big] 
\,.\end{align}

\section{RG Evolution and the Total NLL' Cross Section}
\label{sec:sigmaNLL}

In this section, we apply Renormalization Group (RG) methods to the functions calculated in this paper and arrive at the result for the total NLL' resummed cross section. These functions can be divided into those which are multiplicatively renormalized and those that renormalize via a convolution. The former include the hard function and unmeasured jet functions and the unmeasured part of the soft function, and the latter includes measured jet and soft functions.

\subsection{Hard Function}
\label{ssec:hardRGE}

The hard function $\vect H$ for $N-2$ jet production in hadron collisions is a matrix in color space with rank $R$ (the same as that of the soft function). 
It can be written in terms of Wilson coefficients $C_i$ as $(\vect H)_{ij} = C_i C_j^*$, each of which mix into each other under renormalization, i.e, $C_i^{\rm bare} = \sum_j(\vect{Z}_H(\mu))_{ij} C_j $ which implies that 
\begin{align}
\label{eq:Hrenorm}
 \vect{H}^{\text{bare}} = \vect{Z}_H(\mu)\vect{H}(\mu) \vect{Z}_H^\dagger(\mu)
\,.\end{align}
The $\mu$-independence of the left-hand side of \eq{Hrenorm} implies that $\vect H \equiv \vect{H}(\mu)$ obeys the RGE
\begin{align}
\label{eq:H-RGE}
\frac{d \, \vect H}{d \ln \mu} =  \GammaH \, \vect H + \vect H \,  \GammaH^\dagger
\,,\end{align}
where
\begin{align}
\GammaH \equiv - \vect{Z}_H^{-1} \frac{d}{d\ln \mu} \vect{Z}_H
\end{align}
This RGE preserves the hermiticity of $\vect H$ under RG evolution.
$\GammaH$ in \eq{H-RGE} is given (to $\cO(\as^2)$) by \cite{Chiu:2009mg, Becher:2009qa}
\begin{align}
\label{eq:GammaH}
 \GammaH &= \frac{1}{2}\sum_{i=1}^N \left[\cas{i} \, \Gamma_\cusp(\alpha_s) \ln\frac{m_i^2}{\mu^2} -  \frac{\as}{ \pi} \gamma_i\right] + \Gamma_\cusp(\alpha_s) \,\Mh(m_i)
\,,\end{align}
where $\gamma_i$ is given in \eq{gammaqg}, $\Gamma_\cusp(\as)$ is the cusp anomalous dimension (given in \eq{gamma-gammacusp}),
and $m_i$ is an arbitrary parameter(s) which can be chosen for convenience and can be shown to cancel between the first term and $\Mh(m_i)$. The first term is (implicitly) proportional to an identity matrix and $\Mh$ in the second term involvers a non-trivial matrix of rank $R$, which can be written as
\begin{align}
\label{eq:Mmatrix}
\Mh(m_i) &\equiv - \sum_{\pairs}\Tij \left[ \ln\left((-1)^{\Delta_{ij}}\frac{s_{ij}}{m_i m_j}- i0^+\right)\right] \nn\\
&=  \Ms(m_i) + i \pi \vect T
\,,\end{align}
where $\Delta_{ij}$ is $0$ for beam-jet interference and $1$ for beam-beam and jet-jet interference, $s_{ij} = 2 p_i\mcdot p_j >0$, and in the second line we explicitly separated the terms of the form $\Delta_{ij} \ln(-1)$ into the matrix $i \pi \vect T$,
where
\begin{align}
\label{eq:Tmatrix}
\vect T \equiv \sum_{\pairs}\Delta_{ij} \, \Tij
\,.\end{align}
and  $\Ms(m_i)$ is defined in \eq{SdivD}.
The matrix $\Mh$ is worked out for a set of choices of color bases for all $2 \to 2$ channels in \Ref{Kelley:2010fn} with the choice $m_i^2 = -t > 0$ (the Mandelstam variable) in the $qq' \to qq'$ channel (and the choice for other channels obtained by crossing relations). Importantly, for any $\mu$-independent choice for $m_i$, $\Mh$ is independent of $\mu$. 

The effect of the color-trivial component of \eq{H-RGE} (i.e., the contribution from the term in brackets in \eq{GammaH}) can be obtained using the results in \app{app:RGE} and gives rise to a factor $\Pi_H$ as in \eq{kernel} with the parameters needed for $K_H$ and $\omega_H$ at NLL' given in \tab{anomalous-coeff}. We can straightforwardly include the  effect of $\Gamma_\cusp (\alpha_s) \,\Mh(m_i)$ via matrix exponentiation and record the solution as
\begin{align}
\vect H(\mu, \mu_H) =  \Pi_H(\mu, \mu_H) \, \PiH (\mu, \mu_H) \vect{H}(\mu_H )\PiH^\dagger(\mu, \mu_H)
\,,\end{align}
where
\begin{align}
\label{eq:PiH}
\PiH(\mu, \mu_H)  &\equiv \exp{ \Big\{\Mh \! \int_{\as(\mu_H)}^{\as(\mu)}\frac{d\alpha}{\beta[\alpha]} \Gamma_c(\alpha) \Big\}} = \exp{ \Big\{\Mh\Big( \frac{2}{\beta_0} \ln \frac{\as(\mu_H)}{\as(\mu)}+ \cdots\Big) \Big\}}
\,,\end{align}
where in the second equality we expanded to NLL' accuracy.
This matrix exponential can be defined by first constructing the matrix  $\RH$  of eigenvectors of $\Mh$ such that $\RH^{-1}   \Mh \RH = \vect{\Lambda}_{H}$ is the diagonal matrix of eigenvalues of $\Mh$, and then defining $\exp(\Mh) \equiv \RH \exp(\vect{\Lambda}_{H}) \RH^{-1}$.

\subsection{Jet Functions and Unmeasured Beam Functions}
\label{ssec:jetRGE}

Since the jet functions can be obtained directly from rescalings of those in \Ref{Ellis:2010rwa} as described in \ssec{jet}, the renormalization is similarly related to the results in \Ref{Ellis:2010rwa}. For measured (renormalized) jet functions we have
\begin{align}
\gamma_{J_i}(\tau_a^i, \mu) = \bigg[ 2 \Gamma_{\cusp}(\as) \cas{i} \frac{2-a}{1-a} \ln \frac{\mu}{\pTJ} + \frac{\as}{\pi}\gamma_i \bigg]\delta(\tau_a^i) - 2 \Gamma_{\cusp}(\as) \cas{i} \frac{1}{1-a}\bigg(\frac{1}{\tau_a^i}\bigg)_{\plus}
\,,\end{align}
which is of the general form \eq{gammaFtau}
with cusp ($\Gamma_F[\as]$) and non-cusp ($\gamma_F[\as]$) pieces given in \tab{anomalous-coeff}. Here and below, the `$+$' distribution is defined for example in Eq. (A.2) of \Ref{Ellis:2010rwa}.

To RG evolve the jet function, we perform the integral in \eq{Fconvol} for the case $F=J$. Integrals of this form are most easily performed by convolving the right-hand side against $1 = Z^{-1}\otimes Z$ and first performing the convolution of $U_F$ with the bare function, i.e., $Z\otimes F$, then expanding in $\epsilon$, and finally performing the $Z^{-1}$ convolution (which just removes the $1/\epsilon$ poles in a minimal subtraction scheme). For the jet function, we obtain
\begin{align}
\label{eq:Jresum}
J^{\meas}(\tau_a^i,  \mu) &=   Z_J^{-1}(\tau_a^i, \mu_J) \otimes \Big[J^{\meas}(\tau_a^i) \otimes U_J(\tau_a^i, \mu, \mu_J) \Big] \nn\\
&= Z_J^{-1}(\tau_a^i, \mu_J) \otimes \Bigg\{U_J( \tau_a^i, \mu, \mu_J) \Bigg (1- \frac{\as(\mu_J)}{2 \pi} \bigg[-f_i(a) + \nn\\
& \qquad  \bigg(\frac{1}{\epsilon}\frac{2\cas{i}}{1-a} + \frac{\gamma_i}{1-a/2}\bigg)\frac{\Gamma(-2\epsilon/(2-a))\Gamma(-\omega_J^i)}{\Gamma(-2\epsilon/(2-a) - \omega_J^i)}\bigg(\frac{\mu_J}{\pTJ (\tau_a^i)^{1/(2-a)}}\bigg)^{\!2\epsilon} \bigg] \Bigg) \Bigg\}_{\plus} \nn\\
&=    \bigg\{U_J( \tau_a^i, \mu, \mu_J)  \Big(1+f^i_J(\tau_a^i; \omega_J^i, \mu_J)\Big)\bigg\}_{\plus} 
\,,\end{align}
where  $f_J(\tau_, \Omega, \mu)$ is the one loop part of the renormalized jet function after RG evolution,
\begin{align}
\label{eq:fJ}
f^i_J(\tau, \Omega, \mu) &=  \frac{\as}{ \pi(2-a)} \bigg\{
\frac{2-a}{2}f_i(a) +  \gamma_i\bigg[H(-1-\Omega)+ (2-a) \ln \frac{\mu}{\pTJ \,\tau^{1/(2-a)}} \bigg] \\
& \quad + \frac{\cas{i}}{1-a}\bigg[\bigg(H(-1-\Omega)+ (2-a) \ln \frac{\mu}{\pTJ \,\tau^{1/(2-a)}}\bigg)^2 \nn
- \psi^{(1)}(-\Omega)+\frac{\pi^2}{6}  \bigg]\bigg\}
\,,\end{align}
and $H(x)$ is the harmonic number function and $\psi^{(1)}(x)$ is the polygamma function of order 1 and $f_i(a)$ is given in \eq{fia}.
The natural scale for the jet function suggested by \eq{fJ} is 
\begin{align}
\mu^{\meas}_J  \equiv \pTJ (\tau_a^i)^{1/(2-a)} \, .
\end{align}

From the discussion in \ssec{beam} and the results of \ssec{jet}, we have for both unmeasured jet functions and unmeasured beam functions the anomalous dimensions
\begin{align}
\label{eq:gammaJi}
\gamma_{J_i} = 2 \Gamma_{\cusp}(\as) \cas{i} \ln\frac{\mu}{\pTJ\cR} +\frac{\as}{\pi} \gamma_i
\,,\end{align}
and
\begin{align}
\label{eq:gammaBi}
\gamma_{B_i} = 2 \Gamma_{\cusp}(\as) \cas{i} \ln\frac{\mu}{x_i \Ecm e^{-\yc}} +\frac{\as}{\pi} \gamma_i
\,,\end{align}
which have the form of \eq{gammaF}. We have summarized the cusp and non-cusp parts in \tab{anomalous-coeff} and $\gamma_i$ is given in \eq{gammaqg} for quark and gluon jets. \eqs{gammaJi}{gammaBi} (together with \eq{bareunmeasjet}) suggests the canonical scale choices
\begin{align}
\mu_J^{\unmeas} = \pTJ \cR \qquad  {\rm and } \qquad \mu_B = x_i \Ecm e^{-\yc}
\,,\end{align}
with $x_{i}$ fixed via \eq{x12}.


\subsection{Soft Function}
\label{ssec:softRGE}

The total measured soft function, which includes both the  $\vect{S}^{\rm unmeas}$ and a $S^{\rm meas}$ contribution for each measured jet as in \eq{meas+unmeas}, can be evolved by using a multiplicative-type RGE (cf. \eq{FRGE}) for $\vect{S}^{\rm unmeas}$ and a convolution-type RGE (cf. \eq{RGEtau}) for $S^{\rm meas}$, and each can be evolved from a separate scale (an unmeasured soft scale and a measured soft scale, respectively). This corresponds an early version of ``refactorization'' originally suggested in \Ref{Ellis:2010rwa}. A more complete refactorization procedure was recently introduced in \cite{Chien:2015cka} which  involves further refactorizing $\vect{S}^{\rm unmeas}$ into a global soft contribution and a soft-collinear contribution, as in \eq{Sunmeasrefact}. In this section, we demonstrate how both approaches are achieved so that they can be compared numerically in \ssec{example}.

\subsubsection{Unmeasured Evolution}
\label{sec:SunmeasRGE}

The unmeasured component of the soft function $ \vect{S}^{\unmeas}$ is renormalized much like the hard function\footnote{
Note that \eq{Srenorm} takes the form of \eq{Hrenorm} but with $\vect{Z}_H \leftrightarrow \vect{Z}_S^\dagger$. This gives rise to the RGE \eq{Sunmeas-RGE-1} which is of the form \eq{H-RGE} but with $\vect{\Gamma}_S^{\unmeas} \leftrightarrow \GammaH^\dagger$. RGE invariance then requires $\GammaH = - \vect{\Gamma}_S^\unmeas + \cdots$ where the ellipses denote color-trivial contributions.}
\begin{align}
\label{eq:Srenorm}
 \vect{S}^{\unmeas,\, \text{bare}} = \vect{Z}_S^\dagger(\mu)\, \vect{S}^{\unmeas}(\mu) \, \vect{Z}_S(\mu)
\end{align}
which gives rise to an RGE of the form
\begin{align}
\label{eq:Sunmeas-RGE-1}
 \frac{d}{d\ln\mu}\vect{S}^{\unmeas} = \vect{S}^{\unmeas} \vect{\Gamma}_S^{\unmeas} + \hc
\,,\end{align}
with 
\begin{align}
\label{eq:GammaSunmeas}
\vect{\Gamma}_S^{\unmeas}  &\equiv \frac{\as}{\pi} (\vect S^{\rm div} -   i\pi \vect{T} + \sum_{i=1,2} \cas{i} \ln \cR ) \nn\\
& =\frac{\as}{\pi} \Big( \gammaunmeas(m_i)  - \vect M(m_i)+ \sum_{i=1,2} \cas{i} \ln \cR \Big) 
\,,\end{align}
where $\vect S^{\rm div}$ and $\gammaunmeas$ are defined in \eqs{Sdiv}{SdivD},  and $\vect{M}$ and $\vect{T}$ are defined in \eqs{Mmatrix}{Tmatrix}. In \eq{GammaSunmeas}, we have inserted the factor $ i\pi \vect{T}$ to comply with matrix-level consistency of the anomalous dimensions, which is consistent with the one loop bare soft function calculation \eq{Sunmeas} since  $\vect{S}_0 \vect{T} =\vect{T}^\dagger\vect{S}_0 $.

The solution to this RGE is completely analogous to that of the hard RGE \eq{H-RGE}. The result is
\begin{align}
\label{eq:Sunmeasevolved}
 \vect{{S}}^{\unmeas}(\mu, \mu_S) = \Pi_S^{\unmeas}(\mu, \mu_S) \big[\PiS^\dagger(\mu, \mu_S)  \vect{{S}}^{\unmeas}(\mu_S) \PiS(\mu, \mu_S)\big]
\end{align}
where $\Pi_S^{\unmeas}$ is of the form \eq{kernel} with NLL' parameters given in \tab{anomalous-coeff} and
\begin{align}
\label{eq:PiS}
 \PiS(\mu, \mu_S)  &\equiv \exp{ \Big\{ -\Mh \! \int_{\as(\mu_S)}^{\as(\mu)}\frac{d\alpha}{\beta[\alpha]} \Gamma_c(\alpha) \Big\}} = \exp{ \bigg\{ - \Mh\bigg[ \frac{2}{\beta_0} \ln \frac{\as(\mu_S)}{\as(\mu)}+ \cdots\bigg] \bigg\}}
 \end{align}
where in the second equality we expanded to NLL' accuracy. Inspection of the unmeasured soft function \eq{Sunmeas} suggests the canonical unmeasured soft scale choice
\begin{align}
\mu^{\unmeas}_S \equiv \pTc
\,. \end{align}

\subsubsection{Measured Evolution}
\label{sec:MeasRGE}

When the jets are measured, RGE takes the form
\begin{align}
\label{eq:Smeas-RGE}
\frac{d}{d \ln \mu} \vect S(\tau_a^1, \tau_a^2, \mu)= \int d\tau' d\tau'' [  \vect S(\tau',\tau'', \mu) \, \vect \Gamma_S(\tau_a^1-\tau', \tau_a^2 - \tau'', \mu) + \hc]
\,,\end{align}
with the soft anomalous dimension given to NLL accuracy by
\begin{align}
\label{eq:GammaSmeas}
\vect \Gamma_S(\tau_a^1, \tau_a^2, \mu)  &=\vect{\Gamma}_S^{\unmeas} \delta(\tau_a^1)\delta( \tau_a^2) + \Big[ \frac{1}{2}\gamma_S^{\meas}(\tau_a^1, \mu) \delta(\tau_a^2)  + (1 \leftrightarrow 2) \Big] 
\,,\end{align}
where $\gamma^{\meas}$ is given by
\begin{align}
\gamma_S^{\meas}(\tau_a^i, \mu) = -\Gamma_\cusp(\as) \cas{i} \frac{1}{1-a} \bigg\{ 2 \ln \frac{\mu \cR^{1-a}}{\pTJ}\delta(\tau_a^i) -2 \bigg(\frac{1}{\tau_a^i}\bigg)_{\plus}\bigg\} 
\end{align}
which has the form of \eq{gammaFtau}.
The $\tau_a$ dependence of measured jets requires the inclusion of the  evolution kernels $U^i_S(\tau_a,\mu,\mu_0)$ as in \eq{kernelF} with NLL' parameters given in \tab{anomalous-coeff}. To evaluate the effect of convolving these kernels, we use the same method as in \eqs{Jresum}{fJ}. This gives for the RG evolved measured part of the soft function 
\begin{align}
\label{eq:Sresum}
S^{\meas}(\tau_a^i;  \mu) &=  Z^{-1}_S(\tau_a^i, \mu_S) \otimes \bigg[U^i_S( \tau_a^i, \mu, \mu_S) \bigg(1+\frac{1}{\epsilon} \frac{\as(\mu_S) \cas{i}}{\pi(1-a)} \frac{e^{\gamma_E \epsilon}}{\Gamma(1-\epsilon)} \nn\\
& \qquad \qquad  \qquad \qquad\times \frac{\Gamma(-2\epsilon)\Gamma(-\omega_S^i)}{\Gamma(-2\epsilon - \omega_S^i)}\bigg(\frac{\mu_S \cR^{1-a}}{\pTJ \, \tau_a^i} \bigg)^{\!2\epsilon}\bigg)\bigg]_{\plus} \nn\\
&=  U^i_S( \tau_a^i, \mu, \mu_S) \Big(1+f^i_S(\tau_a^i; \omega_S^i, \mu_S)\Big)
\,,\end{align}
\begin{align}
\label{eq:fS}
f^i_S(\tau; \Omega, \mu) &= \frac{\as \cas{i}}{ \pi(1-a)} \bigg[\psi^{(1)}(-\Omega) - \bigg(H(-1-\Omega) + \ln\frac{\mu \cR^{1-a}}{\pTJ \, \tau}\bigg)^2 - \frac{\pi^2}{8}\bigg]
\,,\end{align}
which suggests the canonical scale choice
\begin{align}
\mu^{\meas}_S \equiv \frac{\pTJ \tau_a^i}{\cR^{1-a}}
\,.\end{align}
Taking the scales from which the two measured components and the unmeasured component are evolved from to be $\muSmeas^{1,2}$ and $\muSunmeas$, respectively, we record the final result as
\begin{align}
\label{eq:Smeasevolved}
\vect S (\tau_a^1, \tau_a^2, \mu, \muSmeas^1, \muSmeas^2, \muSunmeas) &= U^1_S( \tau_a^1, \mu, \muSmeas^1) U^1_S( \tau_a^2, \mu, \muSmeas^2) \big[1 + ( f^1_S(\tau_a^1; \omega_S^1, \muSmeas^1)+f^2_S(\tau_a^2; \omega_S^2, \muSmeas^2)) \big]\nn\\
&  \quad \times
 \Pi_S^{\unmeas}(\mu, \muSunmeas) \big[ \PiS^\dagger (\mu, \muSunmeas)\vect{S}^{\unmeas}(\muSunmeas) \PiS (\mu, \muSunmeas) \big]
\,.\end{align}

\subsubsection{Refactorized Evolution}
\label{sec:RefactorizationRGE}

The components of the refactorized $\vect{S}^\unmeas$ (cf. \eq{Sunmeasrefact}), $\vect{s}_s$ and ${s}_{cs}^k$ for $k=1,2$ evolve as 
\begin{align}
\label{eq:Sunmeas-RGE-2}
 \frac{d}{d\ln\mu}{\vect{s}}_s(E) = \int\! dE' \, \vect{s}_s(E') \,\vect{\Gamma}_{ss}(E-E')
\,,\end{align}
and
\begin{align}
\label{eq:Sunmeas-RGE-3}
 \frac{d}{d\ln\mu}{s}^{k}_{sc}(E \cR) =  \int\! dE' \, s_{sc}^k(E' \cR)  \, \Gamma^k_{sc} ((E-E')\cR)
\,,\end{align}
respectively. The anomalous dimensions take the form \eq{gammaFtau} and satisfy the relations
\begin{align}
\label{eq:gammaSC}
\frac{1}{2}\int_0^{\pTc} \! \!\! d E  \, {\Gamma}^k_{sc}(E) = - \cas{k} \Gamma_{\cusp}[\as] \ln \frac{\mu}{\pTc \cR} + \gamma^k_{\rm hemi}[\as]
\,,\end{align}
and
\begin{align}
\label{eq:gammaSS}
\frac{1}{2}\int_0^{\pTc} \!\!\!d E \, \vect{\Gamma}_{ss}(E) = \sum_{i=1,2} \big( \cas{i} \Gamma_{\cusp}[\as] \ln \frac{\mu}{\pTc}-  \gamma^i_{\rm hemi} [\as] \big) +  \frac{\as}{\pi} ( \gammaunmeas(m_i) - \vect{M}(m_i)) 
\,,\end{align}
where we used that to all-orders, the non-cusp part of the anomalous dimension for $\gamma_{sc}$ is the same as that of the hemisphere thrust distribution \cite{Chien:2015cka} (of the color-representation of jet $k$). At $\cO(\as)$, $\gamma_{\rm hemi}^i = 0$. The additional non-cusp parts of \eq{gammaSS} (which do not appear in the analogous $e^+e^-$ calculation \cite{Chien:2015cka})  are needed for this measurement to ensure the consistency of refactorization at $\cO(\as)$, 
\begin{align}
\label{eq:refactconsistency}
\frac{1}{2}\int_0^{\pTc} \! d E  \,  \big( \vect{\Gamma}_{ss}(E) + \sum_k \Gamma_{sc}^{k}(E) \big) = \vect{\Gamma}^{\unmeas}_S
\,.\end{align}

To RG evolve the refactorized soft function, we write
\begin{align}
\vect{S}_{ss}^{(1)} &= \frac{1}{\epsilon} \bigg( \frac{\mu}{\pTc} \bigg)^{\! 2\epsilon} \vect{f}_s \nn\\
\sum_{k=1,2} S_{sc}^{k(1)} &= \frac{1}{\epsilon} \bigg( \frac{\mu}{\pTc \cR} \bigg)^{\! 2\epsilon} f_c
\end{align}
where $f_{s,c} = \sum_{\{i= 0, 1, 2\}} \epsilon^{i-1} f_{s,c}^{i}$ can be read off from the $\cO(\as)$ results \eqs{Sss1}{Ssc1} and are given by
\begin{align}
{f}_c^{0} &= -2 (\cas{1}+\cas{2} ) 
&
\vect{f}_s^{0} &=  -f_c^0
\nn\\
{f}_c^{1} &=  0 
&
\vect{f}_s^{1} &=  4 \vect{S}^{\rm div}
\nn\\
{f}_c^{2} &=  \frac{\pi^2}{6}(\cas{1}+\cas{2} )
&
\vect{f}_s^{2} &= -8 \vect{T}_1 \cdot \vect{T}_2 \ln(1+e^{\Delta y}) \ln(1+e^{-\Delta y})   - {f}_c^{2}
\,.\end{align} 
This allows us to write the RG evolved bare functions (using a  similar argument as that described above \eq{Jresum}) as
\begin{align}
& \int_0^{\pTc} dE \, \Big[\vect{s}_{s}(E) \otimes U_{ss}(E/\mu_{ss}, \mu, \mu_{ss}) \Big] \otimes_{i=1,2} \Big[{s}_{sc}^i(E \cR)  \otimes {U}_{sc}^i(E \cR/\mu_{sc}, \mu, \mu_{sc})\Big] \nn\\
& =  \int_0^{\pTc} dE \, \Big[\vect{s}_{s}(E) \otimes_{i=1,2} {s}_{sc}^i(E \cR)  \Big] \otimes \Big[ U_{ss}(E/\mu_{ss}, \mu, \mu_{ss}) \otimes_{i=1,2} {U}_{sc}^i(E \cR/\mu_{sc}, \mu, \mu_{sc})\Big] \nn\\
& =  \int_0^{\pTc} dE \, \bigg[1 -2 \frac{\Gamma(-2\epsilon)\Gamma(-\Omega_S)}{\Gamma(-2\epsilon-\Omega_S)}\bigg( \frac{\as(\mu_{ss})}{4\pi}\bigg(\frac{\mu_{ss}}{E} \bigg)^{\! 2 \epsilon}  \vect{f}_s \nn\\
& \qquad \qquad \qquad\qquad\qquad\qquad\qquad\qquad + \frac{\as(\mu_{sc})}{4\pi}  \bigg(\frac{\mu_{sc}}{E \cR} \bigg)^{\! 2 \epsilon} f_c\bigg) \bigg] U_{S}(E, \mu, \mu_{ss}, \mu_{sc})
\,,\end{align}
where in the 3rd line we truncated the series in parenthesis to $\cO(\as)$ and we defined 
\be
\Omega_S \equiv \omega_{ss}(\mu, \mu_{ss}) + \sum_{i=1,2} \omega_{sc}^i(\mu, \mu_{sc}) 
\ee
and
\begin{align}
U_{S}(E, \Omega_S, \mu_{ss}, \mu_{sc}) \equiv \Big[ U_{ss}(E/\mu_{ss}, \mu, \mu_{ss}) \otimes_{i=1,2} {U}_{sc}^i(E \cR/\mu_{sc}, \mu, \mu_{sc})\Big]
\end{align}
and used that $U_S$ scales as
\be
U_S \propto \frac{1}{\Gamma(-\Omega_S)} E^{-1-\Omega_S}
\,.\ee
Expanding in $\epsilon$ and dropping the $1/\epsilon$ poles gives the renormalized, refactorized and RG evolved $\vect{S}^{\unmeas}(\mu)$,
\begin{align}
\label{eq:Sunmeasreplacefullmu}
\vect{S}^{\unmeas}(\mu) 
& \to  \vect{S}^{\unmeas}(\Omega_S, \mu_{ss}, \mu_{sc}) \int_0^{\pTc} \!  dE \, U_{S}(E, \Omega_S, \mu_{ss}, \mu_{sc})
\end{align}
where 
\begin{align}
&\vect{S}^\unmeas(\Omega, \mu_{sc}, \mu_{ss}) \equiv   \vect{S}_0 +\bigg\{ \vect{S}_0\bigg[  \frac{\as(\mu_{ss})}{4\pi} \bigg(\frac{1}{2}\vect{f}_s^2 + 
\vect{f}_s^1 \Big(\ln \frac{\mu_{ss}}{\pTc} + H(-\Omega)\Big) \\
& \qquad\qquad\qquad \qquad\qquad\qquad\qquad+ \vect{f}_s^0 \Big (\frac{\pi^2}{6} - \psi^{(1)}(1-\Omega) +\big (\ln \frac{\mu_{ss}}{\pTc} + H(-\Omega) \big)^2  \Big) \bigg) \nn\\
& \qquad \qquad\qquad\qquad\qquad\qquad\quad+  \frac{\as(\mu_{sc})}{4\pi} \bigg(\frac{1}{2}{f}_c^2 + 
{f}_c^1 \Big(\ln \frac{\mu_{sc}}{\pTc \cR} + H(-\Omega)\Big) \nn\\
& \qquad\qquad\qquad \qquad\qquad\qquad\qquad+ {f}_c^0 \Big (\frac{\pi^2}{6} - \psi^{(1)}(1-\Omega) +\big (\ln \frac{\mu_{sc}}{\pTc \cR} + H(-\Omega) \big)^2  \Big) \bigg)
 \bigg]+ \hc \bigg\} \nn
\,.\end{align}

We note that when combined into the full cross section in \ssec{sigmaRGE}, the $\mu$ dependence can be cancelled to all orders between \eq{Sunmeasreplacefullmu} and the remainder of the cross section (using consistency and \eq{refactconsistency}) at the expense of running all factorized components from $\mu_{ss}$ to the scale of the component. This means for example that we have
\begin{align}
\Omega_S \to \sum_{i=1,2} \omega_{sc}^i(\mu_{ss}, \mu_{sc}) \equiv \omega_{sc}
\,.\end{align} 
This means in particular we can make the replacement
\begin{align}
\label{eq:Sunmeasreplace}
\vect{S}^{\unmeas}(\mu) 
& \to  \vect{S}^{\unmeas}(\omega_{sc}, \mu_{ss}, \mu_{sc}) \, U_{sc}(\omega_{sc}, \mu_{ss}, \mu_{sc})
\end{align}
where
\begin{align}
U_{sc}(\omega_{sc}, \mu_{ss}, \mu_{sc}) \equiv  \int_0^{\pTc}\!  dE \, U_{S}(E,  \omega_{sc}, \mu_{ss}, \mu_{sc})= \frac{e^{K_{sc} + \gamma_E \omega_{sc}}}{\Gamma(1-\omega_{sc})} \bigg( \frac{\mu_{sc}}{\pTc \cR} \bigg)^{\! \omega_{sc}}
\,,\end{align}
where $K_{sc} \equiv \sum_{i=1,2} K_{sc}(\mu_{ss}, \mu_{sc})$. The parameters needed for $K_{sc}$ and $\omega_{sc}$ at NLL'  (which can be expanded as in \eq{kernelparamsNLL}) can be read off from \eqs{gammaSC}{gammaSS} and are given in  \tab{anomalous-coeff}.

\TABLE[t]{
\begin{tabular}{ | c || c | @{}c | c | c | c |}
\hline
 & $ \Gamma_F[\alpha_s] $ & $ \gamma_F[\alpha_s] $ & $j_F$ & $m_F$ & $\mu_F$\\
\hline
\hline
$ \gamma_H $ & \tabrule $-\Gamma \sum_i \cas{i}$ &  $-\sum_i \frac{\as}{\pi} \gamma_i  $ & 1 & $\prod_i m_i^{\cas{i}/\sum_j \cas{j}}$ & $m_i$\\
\hline
$\gamma_{J_{i}}(\tau_a^i)$ &  \tabrule  $\Gamma  \cas{i} \frac{2-a}{1-a}$ & $\frac{\as}{\pi}\gamma_i $  & $2-a$ & $\pTJ$ & $ \pTJ (\tau_a^i)^{1/(2-a)}$ \\
$\gamma_S^\meas(\tau_a^i)$\!\! & \tabrule $  - \Gamma\cas{i} \frac{1}{1-a}$ & $0$  & 1 & $\pTJ/\cR^{1-a}$ & $\pTJ \, \tau_a^i/\cR^{1-a}$ \\
\hline
$\gamma_{J_{i}}$ & \tabrule $\Gamma \cas{i}$  & $\frac{\as}{\pi}\gamma_i$  & $1$ & $\pTJ \cR$ & $\pTJ \cR$ \\
\hline
$\gamma_{B_i}$ & \tabrule $\Gamma \cas{i}$  & $\frac{\as}{\pi}\gamma_i$  & $1$ & $x_i\Ecm e^{-\yc}$ &  $x_i\Ecm e^{-\yc}$ \\
\hline
\rule{0pt}{4ex}
\rule[-3ex]{0pt}{0pt}
 $\gamma_S^\unmeas$ & \tabrule $0$ & \, \begin{tabular}{@{}c@{}}$\frac{2 \as}{\pi}\gammaunmeas(m_i)$  \\ $+ \frac{2 \as}{\pi} (\cas{1} + \cas{2}) \ln \cR $\end{tabular}   & 1 & --- & $\pTc$  \\
\hline
$\gamma_{ss}$ & \tabrule $ \Gamma (\cas{1}+ \cas{2})$ & \, $\frac{2 \as}{\pi}\gammaunmeas(m_i)$  & 1 & $\pTc$  & $\pTc$ \\
\hline
$\gamma_{sc}^i$ & \tabrule $ -\Gamma\cas{i}$ & \, $ 0 $  & 1 & $\pTc \cR$ & $\pTc \cR$ \\
\hline
\end{tabular}
\caption{
Ingredients for anomalous dimensions of the color-trivial parts components to the factorization formula and the corresponding canonical scale choices $\mu_F$, which take the form of \eqs{gammaF}{gammaFtau}. The hard and (unmeasured) soft components require an additional color-nontrivial factor derived explicitly in the text. Here, $\cas{i}$ is the quadratic Casimir ($\CF$ or $\CA$ for quarks and gluons, respectively), $\gamma_i$ is given in \eq{gammaqg}, $\Gamma \equiv \Gamma_\cusp(\as)$ is the cusp (given in \eq{gamma-gammacusp}), $x_i$ are the momentum fractions of the partons in the beams (fixed via \eq{x12}), and $\gammaunmeas$ is given in \eq{SdivD} (and $m_i$ is an arbitrary parameter that cancels both within $ \GammaH$ and within $\vect \Gamma_S$ and can for example be chosen based on the partonic channel to coincide with the conventions of \Ref{Kelley:2010fn} as described in the text). For refactorizing the soft function as in \cite{Chien:2015cka}, the last two rows are used in place of $\gamma_S^\unmeas$.
}
\label{tab:anomalous-coeff}
}

\subsection{Total NLL' Resummed Cross Section }
\label{ssec:sigmaRGE}

For the case of unmeasured jets, we can now readily assemble the ingredients in \eq{sigmafactunmeas} to obtain
\begin{align}
\label{eq:finalunmeas}
d\sigma &= \Norm  \Beam (x_1, \mu_B^1) \Beambar(x_2, \mu_B^2)  J_1(\muJunmeas^1) J_2(\muJunmeas^2) \, \Pi^{\unmeas}(\muSunmeas, \muJunmeas^{1,2}, \mu_B^{1,2},  \mu_H) 
 \nn\\ & \qquad \times
  \Tr \{ \vect H(\mu_H) \vect{\Pi}^\dagger (\muSunmeas, \mu_H) \vect S^{\unmeas} (\muSunmeas) \vect{\Pi}(\muSunmeas, \mu_H)\}
\end{align}
where here and below we 
use a bar over a parameter to denote that it is an unmeasured quantity (so for example $\muSunmeas$ denotes the unmeasured soft scale while $\muSmeas$ denotes the measured soft scale), and $x_{1,2}$ are fixed to the values in \eq{x12}. The function $\Pi$ in \eq{finalunmeas} is defined as
\begin{align}
\vect{\Pi}(\muSunmeas, \mu_H) = \PiS(\mu, \muSunmeas)  \PiH(\mu, \mu_H) 
&= \exp{ \Big\{ \Mh \! \int_{\as(\mu_H)}^{\as(\muSunmeas)}\frac{d\alpha}{\beta[\alpha]} \Gamma_c(\alpha) \Big\}} 
\nn\\&
 =   \exp{ \bigg\{  \Mh \bigg[ \frac{2}{\beta_0} \ln \frac{\as(\mu_H)}{\as(\muSunmeas)}+ \cdots\bigg] \bigg\}}
\end{align}
with $\PiH$ and $\PiS$ defined in \eqs{PiH}{PiS}, respectively, where in the second equality we canceled the $\mu$ dependence (to all orders) and in the third equality we expanded to NLL' accuracy. We also used the definition of the overall multiplicative RG kernel as
\begin{align}
\label{eq:Piunmeas}
 \Pi^{\unmeas}(\muSunmeas, \muJunmeas^{1,2}, \mu_B^{1,2},  \mu_H)  &\equiv  \Pi_H(\mu, \mu_H) \Pi_S^{\unmeas}(\mu, \muSunmeas) \prod_{i=1,2} \Pi^i_B(\mu, \mu_B^i) \prod_{i=1,2}\bar{\Pi}^i_J(\mu, \muJunmeas^i) \nn\\
 & = \prod_{F = H,\Beam_1, \Beam_2, J_1, J_2} e^{K_F(\muSunmeas, \mu_F)} \bigg(\frac{\mu_F}{m_F} \bigg)^{\omega_F(\muSunmeas, \mu_F)}
\,,\end{align}
where $m_F, K^i_F, \omega^i_F$ for $F=J_i,B_i,H$ are given to NLL' in \eq{kernelparamsNLL} in terms of the parameters of \tab{anomalous-coeff}. To arrive at \eq{Piunmeas}, we used the consistency of the anomalous dimensions to explicitly cancel the $\mu$ dependence to all orders.
Here and below, we denote unmeasured quantities with bars to distinguish them from the corresponding measured quantities below.

When the angularity of one or more jets is measured, we need to include $S^{\meas}(\tau_a^i)$ (and its corresponding anomalous dimension $\gamma_S^\meas(\tau_a^i)$) for each measured jet, and we need to replace the unmeasured jet functions $J_i$ with measured ones $J(\tau_a^i)$ (and replace $\bar{\Pi}_{J}^i \to U_J(\tau_a^i)$).
To perform the convolutions for measured jet functions with the measured part of the soft functions, it is easier to first do the convolutions of the evolution factors with each other, and then convolve the resulting full kernel with the renormalized functions. For the case of two measured jets, this yields
\begin{align}
\label{eq:finalmeas}
d\sigma(\tau_a^1 ,\tau_a^2) &= 
 \Norm  \Beam (x_1, \mu_B^1) \Beambar(x_2, \mu_B^2)  \bigg[\Pi^{\meas}(\tau_a^{1,2}, \muSmeas^{1,2}, \muSunmeas, \muJmeas^{1,2}, \mu_B^{1,2}, \mu_H)  \nn\\
 & \qquad \qquad\qquad\times\big[1+ \big( f^1_S(\tau_a^1; \omega_S^1, \mu_S^1)+f^1_J(\tau_a^1; \omega_S^1, \mu_J^1) + (1 \leftrightarrow 2)\big) \big] \bigg]_{\plus} \nn\\ 
 &  \qquad \times \Tr \Big\{ \vect H(\mu_H) \vect{\Pi}^\dagger (\muSunmeas, \mu_H) \vect S^{\unmeas}(\muSunmeas)  \vect{\Pi}(\muSunmeas, \mu_H) \Big\}  
\,,\end{align}
where $f_J^i(\tau,\Omega,\mu)$ and $f_S^i(\tau,\Omega,\mu)$ are given in \eqs{fJ}{fS}, respectively, and
we defined
\begin{align}
\label{eq:Pimeas}
&\Pi^{\meas}(\tau_a^{1,2}, \muSmeas^{1,2}, \muSunmeas, \muJmeas^{1,2}, \mu_B^{1,2}, \mu_H)  \nn\\
&\qquad \qquad  \equiv \frac{ \Pi^{\unmeas}(\muSunmeas, \muJmeas^{1,2}, \mu_B^{1,2},  \mu_H) }{ \prod_{i=1,2}\bar{\Pi}^i_J(\mu, \muJmeas^i) }\! \prod_{i=1,2} \! U_J^i(\tau_a^i, \mu, \mu_J^i)  \otimes  U_S^i(\tau_a^i, \mu, \mu_S^i)  \nn\\
 & \qquad \qquad =\Pi^{\unmeas}(\muSunmeas, \muJmeas^{1,2}, \mu_B^{1,2},  \mu_H) \prod_{i=1,2}  \frac{e^{ K_S^{i} + \gamma_E \, \omega_S^i }}{\Gamma(-\omega_S^i)} \left(\frac{\muSmeas^i}{m_S^i} \right)^ {\! \omega_S^i} \frac{\Theta(\tau_a^i)}{(\tau_a^i)^{1+\omega_S^i}}
\,,\end{align}
where $\gamma_E$ is the Euler constant. The $K_S^i$ and $\omega_S^i$ appearing in these \eqs{finalmeas}{Pimeas} are expanded to NLL' in \eq{kernelparamsNLL} in terms of the parameters in \tab{anomalous-coeff} and are evaluated at the scales
\begin{align}
\omega_S^i &\equiv \omega_S^i(\muJmeas^i, \muSmeas^i) \nn\\
K_S^i &\equiv K_S^i(\muJmeas^i, \muSmeas^i)
\,.\end{align}
To arrive at \eq{Pimeas}, we used that
\begin{align}
\label{eq:meas-unmeas-consistency}
\gamma_{J_i}(\tau_a^i, \mu) + \gamma_S^{\meas}(\tau_a^i, \mu) - \gamma_{J_i}(\mu) \, \delta(\tau_a^i) = 0
\end{align}
to explicitly cancel the $\mu$ dependence of the measured jet and soft functions and the subtracted out unmeasured jet functions (evaluated at the measured jet scale $\muJmeas$). In particular, \eq{meas-unmeas-consistency} implies that 
\begin{align}
e^{K^i_S(\muJmeas, \muSmeas)}\bigg(  \frac{\muSmeas}{m_S}\bigg)^{j_S \omega^i_S(\muJmeas, \muSmeas)} &= e^{K^i_J(\mu, \muJmeas)+K^i_S(\mu, \muSmeas)- \bar{K}^i_J(\mu, \muJmeas)} \\
& \quad \times\bigg(  \frac{\muJmeas}{\mJmeas}\bigg)^{j_J \omega^i_J(\mu, \muJmeas)}\bigg(  \frac{\muSmeas}{m_S}\bigg)^{j_S \omega^i_S(\mu, \muSmeas)} \bigg(  \frac{\muJmeas}{\mJunmeas}\bigg)^{- \bar{\omega}^i_J(\mu, \muJmeas)}  \nn
\,,\end{align}
and that
\begin{align}
\omega^i_S(\muJmeas, \muSmeas) = \omega^i_S(\mu, \muSmeas) +  \omega^i_J(\mu, \muJmeas) 
\,.\end{align}

Finally, we note that to refactorize the cross section and resum logarithms of $\cR$ as in \Ref{Chien:2015cka}, we simply need to make the replacement \eq{Sunmeasreplace} for both the case of unmeasured and of measured jet formula, \eqs{finalunmeas}{finalmeas}, respectively, and interpret $\muSunmeas \to \mu_{ss}$. We discuss the numerical impact of this effect in the next Section.

\subsection{A Simple Example}
\label{ssec:example}

We consider the simple partonic channel $q q' \to q {q}'$.  Of course to compute a physically observable cross section we will need to sum over all partonic channels, however, this is beyond the scope of this work. 
Our aim is to consider the scale variation of the cross section and investigate the impact of refactorization of the soft function on the differential cross section.  We find the main effect of refactorization is to reduce the normalization of the cross section and to lower the scale uncertainty, which is qualitatively similar to what is found in the study of refactorization in $e^+e^-$ collisions recently completed in \Ref{Chien:2015cka}. We also study the dependence of the cross section on the parameters $\cR$, $p_T^{\rm cut}$, and $a$, and comment on the physics responsible for this dependence.

From the results of \Ref{Kelley:2010fn} we have the ($\overline {\rm MS}$ renormalized) hard function to $\cO(\as)$ in the color basis that corresponds to the t-channel $\mathbb 8 \otimes \mathbb 8$ and $\mathbb{1} \otimes \mathbb 1$ operators, 
\begin{align}
\vect H(\mu) = 8 g^4 \big(\vect H_0 + \frac{\as}{4\pi}\vect H_1 (\mu) + \cO(\as^2) \big)
\,,\end{align}
where
\begin{align}
\vect H_0 = \frac{s^2+u^2}{t^2}
\begin{pmatrix}
1 &  0\\
0 & 0
\end{pmatrix}
\,,\end{align}
and
\begin{align}
[\vect H_1(\mu)]_{11} &= \frac{s^2+u^2}{t^2} \bigg(-4 \CF \ln^2\frac{-t}{\mu^2} +2 \Real [  X_1(s,t,u) ]\ln\frac{-t}{\mu^2}  +2Y \bigg) \nn\\
& \qquad + \frac{s^2}{t^2} \Big(\CA - 4 \CF \Big) \Real[Z(s,t,u)]  + \frac{u^2}{t^2} (4 \CF -2 \CA) \Real[Z(u,t,s)] \nn\\
[\vect H_1(\mu)]_{21} &=   \frac{s^2+u^2}{t^2} X_2(s,t,u) \ln \frac{-t}{\mu^2} -  \frac{s^2}{t^2} \frac{\CF}{2\CA}Z(s,t,u) +  \frac{u^2}{t^2} \frac{\CF}{2\CA} Z(u,t,s)\nn\\
[\vect H_1(\mu)]_{12} &=[\vect H_1(\mu)]_{21}^* \nn\\
[\vect H_1(\mu)]_{22} &= 0
\,,\end{align}
where $X_{1,2}$, $Z$, and $Y$ are defined in Eqs.~(33)-(36) of \cite{Kelley:2010fn} and $s$, $t$, and $u$ are given in terms of the jet rapidities and $\pTJ$ in \eq{mandelstam}.

To use the convention of \cite{Kelley:2010fn}, we set $m_i = \sqrt{-t}$ for this channel and have
\begin{align}
\vect \Ms(\sqrt{-t}) = 
\begin{pmatrix}
4 \CF \ln \frac{-u}{s} - \CA \ln\frac{t u}{s^2} \quad & 2 \ln \frac{-u}{s}\\
\frac{\CF}{\CA}\ln \frac{-u}{s} & 0 
\end{pmatrix}
\end{align}
and 
\begin{align}
\vect \Mh(\sqrt{-t}) = \vect \Ms(\sqrt{-t}) + i \pi \vect T\, ,
\end{align}
where
\begin{align}
\vect T = 
\begin{pmatrix}
-2/\CA &  2\\
 \CF/\CA & 0
\end{pmatrix}
\,.\end{align}
Computing the eigenvalues of $\vect M$ gives 
\begin{align}
\lambda^H_{1,2} &= -\frac{\CA}{2} \Big(\ln \frac{ut}{s^2} + 2 i \pi\Big) + 2 \CF \Big(\ln \frac{-u}{s} + i \pi\Big)\nn\\
& \qquad \pm \sqrt{\frac{\CA^2}{4} \Big(\ln\frac{ut}{s^2}   + 2 i \pi\Big)^2  - 2 \CF\CA \Big(\ln \frac{-u}{s} + i\pi\Big) \Big(\ln \frac{-t}{s}+i \pi\Big)}
\,,\end{align}
and for the eigenvectors we find
\begin{align}
\RH &= 
\begin{pmatrix}
\lambda^H_1 &  \lambda^H_2\\
 \frac{\CF}{\CA} \Big( \ln \frac{-u}{s} + i\pi\Big) &  \frac{\CF}{\CA}  \Big( \ln \frac{-u}{s} + i\pi\Big) 
\end{pmatrix}
\,.\end{align}

The $\overline {\rm MS}$ renormalized soft function for the naive factorization is given by
\begin{align}
\vect S^{\unmeas}(\mu) &=\vect S_0 +\frac{\as }{\pi} \bigg\{ \vect S_0 \bigg[ ( \vect S^{\rm div} + 2 \CF \ln \cR )\ln \frac{\mu}{\pTc}- \CF \log^2 \cR 
\nn\\& \qquad  \qquad \qquad 
-  \vect T_1 \mcdot \vect T_2 \log\big(1+ e^{\Delta y}\big)\log\big(1+ e^{-\Delta y}\big)\bigg] + \hc \bigg\}
\,,\end{align}
whereas the refactorized result is obtained with the replacement \eq{Sunmeasreplace}.
The tree level soft function in this basis is given by 
\begin{align}
\vect S_0 = 
\begin{pmatrix}
\frac{1}{2} \CF \CA &  0\\
0 & \CA^2
\end{pmatrix}
\end{align}
In addition to $S_0$ and the matrix component $\vect M'(m_i)$ of $\vect S^{\rm div}$ given above, we need the matrix $\vect T_1 \mcdot \vect T_2$, which for a general $2 \to 2$ scattering is given by
\begin{align}
\vect T_1 \mcdot \vect T_2 = \vect T_{B} \mcdot \vect T_{\Bbar} + \frac{1}{2}(\cas{B} + \cas{\Bbar} -\cas{1} - \cas{2} )
\,.\end{align}
For $qq \to qq$,  $\cas{i} = \CF$ for all $i$ so the $C_i$ cancel and we have
\begin{align}
\vect T &= \frac{1}{2} \Big[ 2 \vect T_{B} \mcdot \vect T_{\Bbar} + 2 \vect T_1 \mcdot \vect T_2 \Big] = 2 \vect T_1 \mcdot \vect T_2 \, .
\end{align}

\FIGURE[t]{
\includegraphics[width = \textwidth]{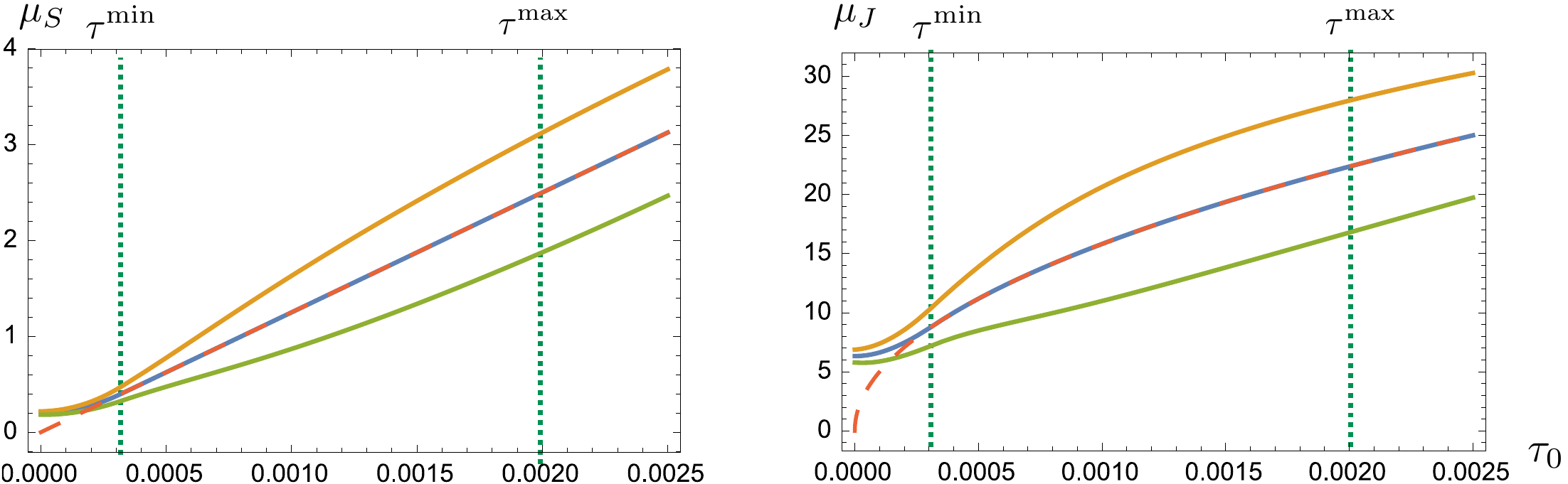}

{ \caption[1] {Profile functions for $\mu_S$ and $\mu_J$. These functions are defined in Eq.~(\ref{def}) and below.}
\label{fig:profiles}}
}

To estimate uncertainty from higher orders in perturbation theory, we vary the hard scale $\mu_H$ and the unmeasured jet and soft scales, $\muJunmeas$ and $\muSunmeas$, separately by $\pm50\%$ around their central values, which we take to be the canonical scales $\mu_F$ given in \tab{anomalous-coeff}. For the refactorized case, we vary the soft scales $\mu_{ss}$ and $\mu_{sc}$ simultaneously. However, to avoid varying the measured jet and soft scales for $\mu_{J,S} \sim \Lambda_{\rm QCD}$, we vary them around profile functions \cite{Ligeti:2008ac, Abbate:2010xh}. This is done by defining $\mu_{J,S}$ as 
\begin{align}\label{def}
\muSmeas^i(\tau_a^i) &= (1+ e_S g(\tau)) \mu(\tau_a^i) \nn\\
\muJmeas^i(\tau_a^i) &= (1+ e_J g(\tau))\big(\pTJ \cR \big)^{\frac{1-a}{2-a}} \big( \mu(\tau_a^i)\big)^{\frac{1}{2-a}}
\,.\end{align}
with $e_{J,S} \in (-1/2, 1/2)$.  The total uncertainty bands are defined to be the envelope of all of the above variations.

In terms of the function
\begin{align}
\theta_\epsilon(x) \equiv \frac{1}{1+ \exp{(-x/\epsilon)}}
\,,\end{align}
which becomes a Heaviside step function in the limit $\epsilon \to 0$,
\begin{align}
\lim_{\epsilon \to 0} \theta_\epsilon(x) = \theta(x)
\,,\end{align}
the function $g(\tau)$ is chosen to be
\begin{align}
\label{eq:gtau}
g(\tau) = \theta_{\epsilon_1}\!(\tau-\tau^{\rm min}) \, \theta_{\epsilon_2}\!(\tau^{\rm max}-\tau)
\,,\end{align} 
and $\mu(\tau)$ is chosen to be
\begin{align}
\label{eq:profile}
\mu(\tau) = 
\begin{cases}
&\mu_0 + \alpha \tau^\beta \sqrt{-t}, \qquad \, \tau<\tau^{\rm min} \\
& \dfrac{\pTJ \, \tau}{\cR^{1-a}}, \qquad   \,\, \, \tau>\tau^{\rm min} \,, 
\end{cases}
\end{align}
where $\alpha$ and $\beta$ are fixed by the continuity of $\mu(\tau)$ and its first derivative to be
\begin{align}
\alpha &= \frac{\pTJ}{\beta (\tau^{\rm min})^{\beta-1} \cR^{1-a} \sqrt{-t}} \nn\\
\beta &= \bigg(1 - \frac{\mu_0 R^{1-a}}{ \pTJ \tau^{\rm min}} \bigg)^{\!-1} 
\,,\end{align}
respectively. The continuity conditions also require that $\beta$ is  greater than unity which implies we need $ \tau^{\rm min} >  \mu_0 \cR^{1-a}/\pTJ $.

\FIGURE[t]{
\includegraphics[width = \textwidth]{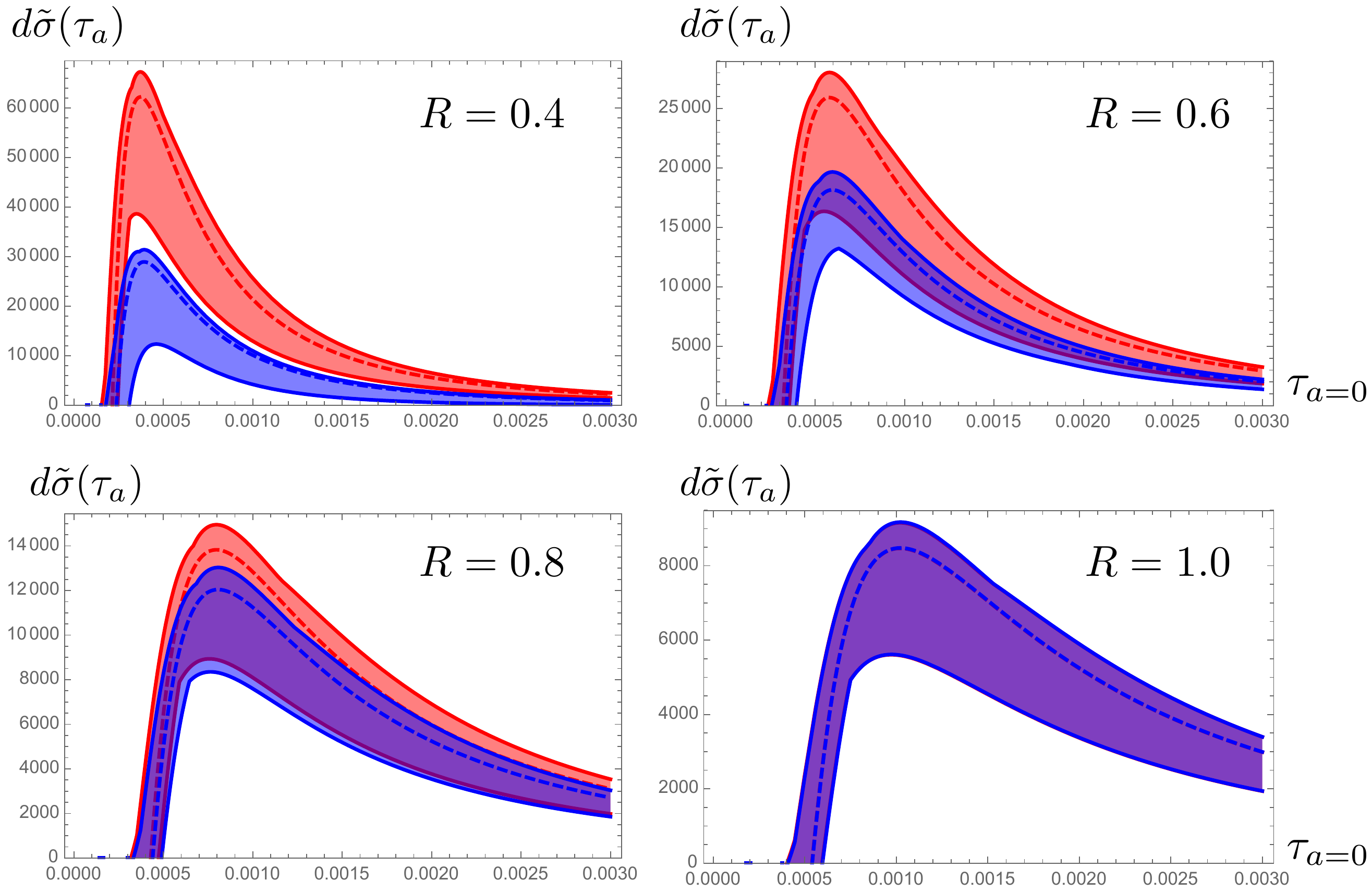}

{ \caption[1] {Differential cross section for four different values of $R$ with soft function refactorized  (blue) and without (red). Central values are dotted lines and band includes scale variation.}
\label{FactRefactvsR}}
}

\FIGURE[t]{
\includegraphics[width = \textwidth]{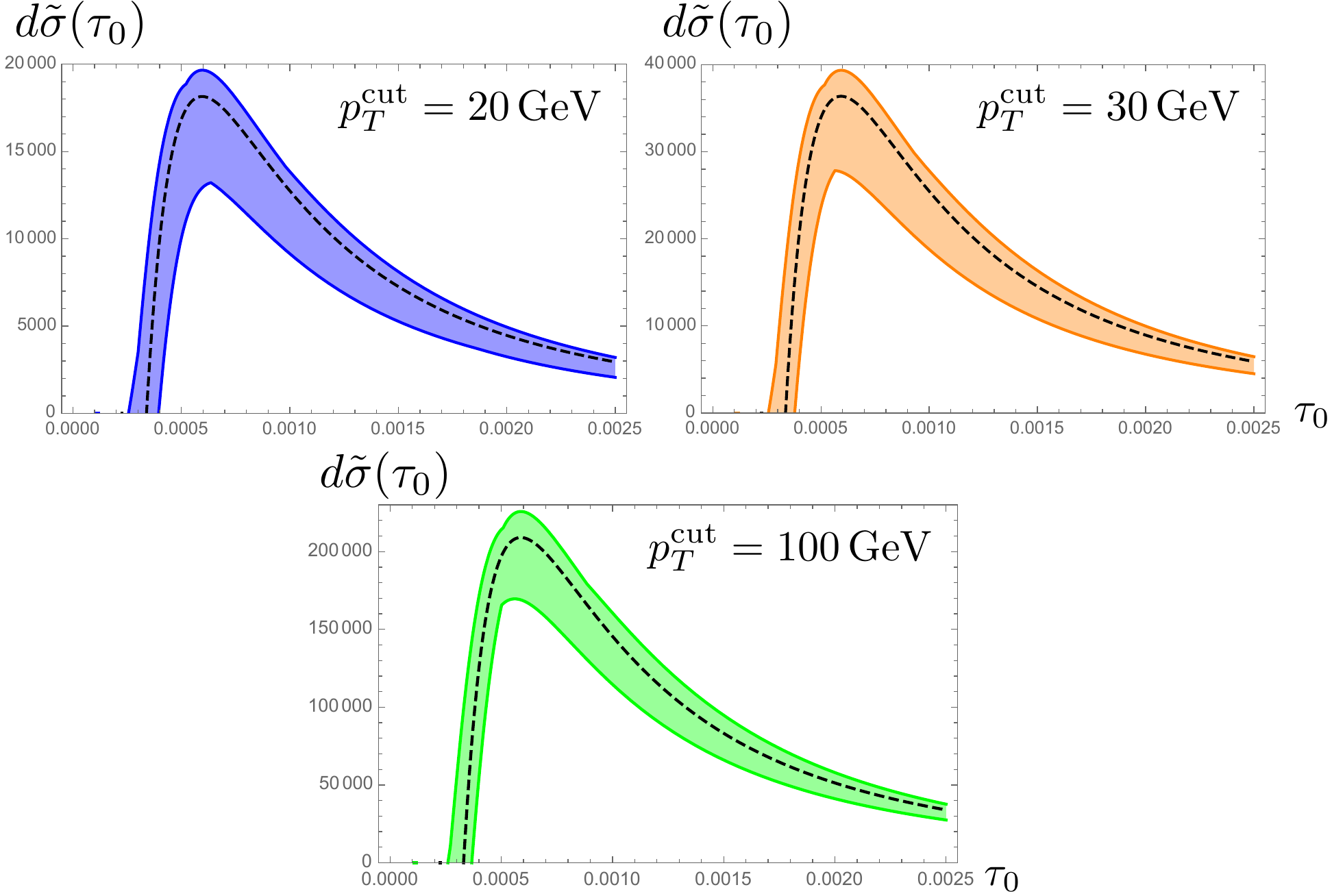}

{ \caption[1] {Differential cross sections for three different values of $p_T^{\rm cut}$.}
\label{fig:pTcutdependence}}
}

The profile functions for $\mu_S$ and $\mu_J$, for $a=0$, are shown in Fig.~\ref{fig:profiles}.
\eqs{gtau}{profile} together ensure that for sufficiently small $\tau$, the scale choice becomes frozen to be $\mu_0$ (and non-perturbative physics dominates), above some scale $\tau^{\rm min}$ we recover the canonical choices (cf. $m_{J,S}$ of table \tab{anomalous-coeff}), and above a third scale $\tau^{\rm max}$ individual $H,J,S$ scale variation begins to dampen (as that should be handled by the traditional $\mu$ variation of fixed-order QCD using a tail-region matching scheme). This is expected to give reasonable scale variation for the range of validity, roughly $\tau^{\rm min} < \tau < \tau^{\rm max}$.

For the sake of illustration, we plot the ``normalized cross section'' (which neglects the PDFs and effects of the fixed order beam function corrections, the latter of which can be found in \cite{Procura:2011aq} following the discussion in \ssec{beam}), defined as
\begin{align}
\label{eq:sigmatilde}
d\tilde{\sigma}(\tau_a) \equiv \frac{B(x_1, \mu= \mu_H)\Bbar(x_2, \mu= \mu_H)}{B(x_1, \mu= \mu_B^1)\Bbar(x_2, \mu= \mu_B^2)} \frac{d\sigma(\tau_a^1, \tau_a^2)}{\sigma^{\rm LO}(\mu = \mu_H)} \bigg \vert_{\tau_a^1 = \tau_a^2 = \tau_a}
\,.\end{align}
For the kinematic and algorithm/observable parameters, we choose for a set of default parameters (fixed to these values unless explicitly varying them in the figures)
\begin{equation}
\label{eq:kinematicnumbers}
\begin{aligned}
\qquad \Ecm &= 10 \TeV \\
a&=0
\end{aligned}
\qquad 
\begin{aligned}
y_1&=1.0\\
y_2&=1.4
\end{aligned}
\qquad
\begin{aligned}
p_T&=500 \GeV\\
\pTc&=20 \GeV
\end{aligned}
\qquad
\begin{aligned}
\cR &=0.6 \\
\yc&=5.0
\end{aligned}
\,,\end{equation}
which corresponds to (via \eqs{x12}{mandelstam})
\begin{align}
\label{eq:stunumbers}
\begin{aligned}
t/s&= - 0.401\\
u/s&= - 0.599\\
\sqrt{s}/\Ecm&= 0.051 
\end{aligned}
\qquad {\rm and } \qquad 
\begin{aligned}
x_1 &= 0.169\\
x_2 &= 0.015
\end{aligned}
\,,\end{align}
and for the profile functions parameters, we choose
\begin{align}
\label{eq:profilenumbers}
\begin{aligned}
\tau^{\rm min} &= 2 (1-a) \mu_0 {\cal R}^{1-a}/p_T = .00032 (1-a)\\
\tau^{\rm max} &= .002
\end{aligned}
\qquad {\rm and } \qquad 
\begin{aligned}
\frac{\epsilon_{1}}{\tau^{\rm min}} &= \frac{\epsilon_2}{\tau^{\rm max}} = 10^{-0.1} \\
\mu_0 &= 200 \, {\rm MeV}
\end{aligned} 
\,.\end{align}

In Fig.~\ref{FactRefactvsR} we show the NLL' calculations for four different values of $\cR$, with all other parameters set to their default values in Eq.~(\ref{eq:kinematicnumbers}). In these plots the blue bands are the predictions with a refactorized soft function and the red bands are the predictions without refactorization. In the limit $\cR \to 1$ the scales $\mu_{ss}$ and $\mu_{sc}$ coincide and the two calculations must give the same result, as seen in the figure. For the smallest value of $\cR=0.4$,  refactorization lowers the normalization of the cross sections by a factor of roughly two, without changing the shape of the distribution or the location of the peak. Refactorization gives a small reduction in the scale uncertainty for $\cR <1$. Note that as $\cR$ decreases the peak in the $\tau_0$ distribution shifts to smaller values of $\tau_0$ because the jets are narrower.
 
 Fig.~\ref{fig:pTcutdependence} shows the refactorized NLL' resummed cross section for three different values of $p_T^{\rm cut}$ with all other parameters set to their defaults in Eq.~(\ref{eq:kinematicnumbers}). Interestingly the shape of the distribution and the location of the peak in the cross section are completely independent of $p_T^{\rm cut}$, only the normalization of the cross section is affected. As expected, the cross section is larger for larger values of $p_T^{\rm cut}$. As discussed in the Introduction, the NGLs, which are of the form $\as^n \ln^n (\pTc \, \cR^2/p_T^J \, \tau_a)$, for $n\geq 2$, combine $p_T^{\rm cut}$ and $\tau_a$  in a nontrivial way. It is possible that when the NGLs are included in the calulcation, the 
 location of the peak of the $\tau_a$ distribution may no longer be $p_T^{\rm cut}$ independent.  Therefore, the dependence of the peak on $p_T^{\rm cut}$ might be an observable that is sensitive to the NGLs.

\FIGURE[t]{
\includegraphics[width = \textwidth]{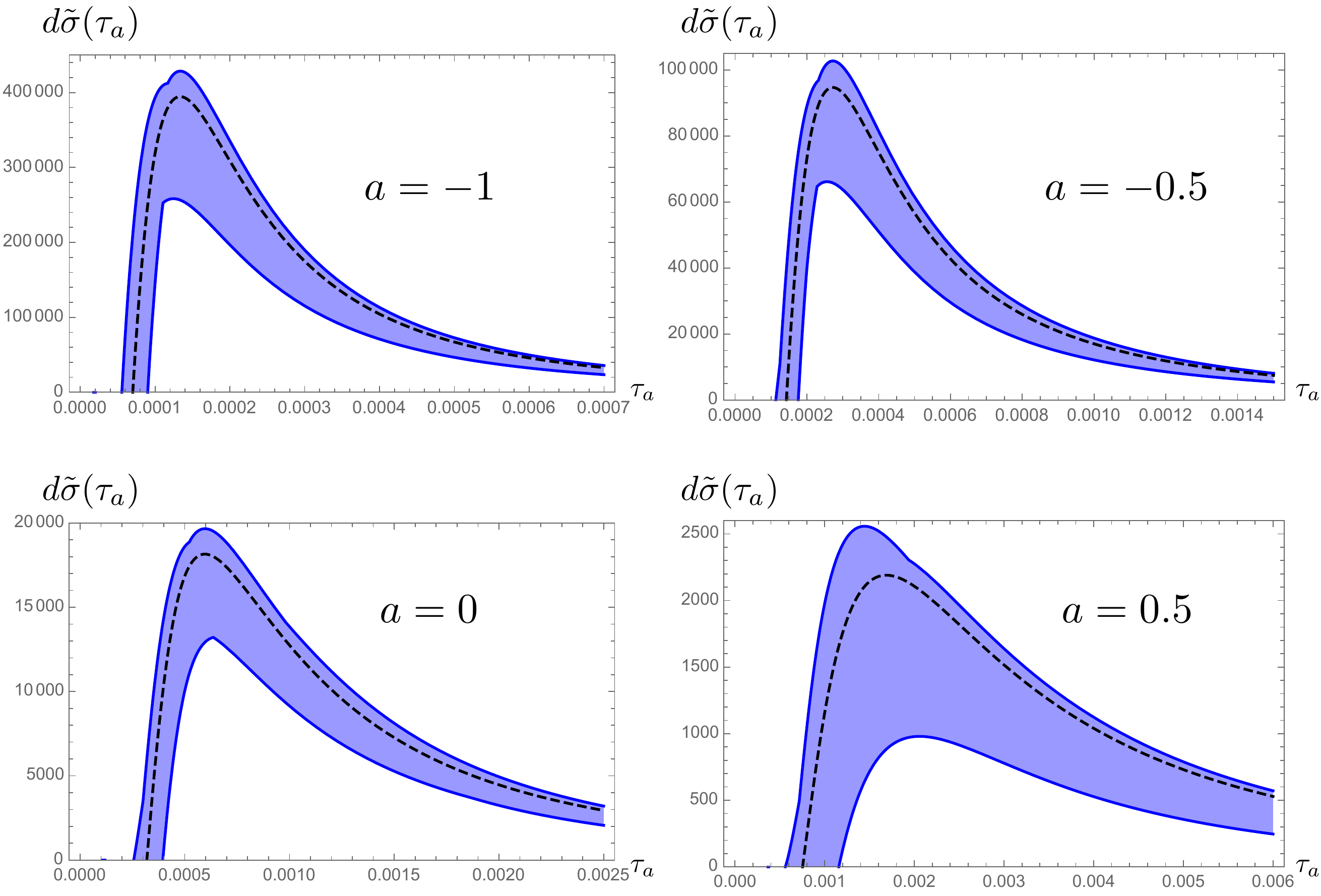}

{ \caption[1] {Differential cross sections for four different values of $a$.}
\label{fig:adependence}}
}

Fig.~\ref{fig:adependence} shows the refactorized NLL' resummed cross section for four different values of $a$ with other parameters set to the default values.  
As $a$ is made large and negative, the contribution to the angularity from particles collinear to the jet axis is suppressed by large powers of the angle with the jet axis.
Correspondingly the distribution is peaked at smaller values of $\tau_a$, a behavior  also seen in calculations of jet angularities in $e^+e^-$ collisions~\cite{Ellis:2010rwa}. 
It is important for obtaining sensible scale variation for all values of $a$ that the parameter $\tau^{\rm min}$ defined in Eq.~(\ref{eq:profilenumbers}) is proportional $(1-a)$.
Both perturbative and power corrections grow with $1/(1-a)$ and factorization 
breaks down completely for $a=1$ in ${\rm SCET}_{\rm I}$ (although an ${\rm SCET}_{\rm II}$ approach can be used for $a=1$ \cite{Becher:2011pf,Chiu:2012ir}). Thus, one expects increasing uncertainty as $a\to1$ from below, and we see from Fig.~\ref{fig:pTcutdependence} that the uncertainties in the 
predictions are substantially larger for $a=0.5$ than for $a \leq 0$.


\section{Conclusion}
\label{sec:conclusion}

In this work, we presented the factorization formulae valid for jet production in hadron colliders with rapidity cuts about the beams, an out-of-jet $\pTc$ veto, and the jets identified with either a $k_T$-type (including $k_T$, $C/A$, and anti-$k_T$) or cone-type algorithm. We considered the cases that the jets can either be identified but otherwise unprobed (``unmeasured'' jets) or are further probed with angularities (``measured'' jets). The ingredients of these formulae involved jet functions, unmeasured beam functions, and an observable dependent soft function. This soft function was further written in terms of a universal piece, $\vect{S}^{\rm unmeas}$,  which encodes the out-of-jet energy veto $\pTc$ and angularity independent (but color and direction dependent) pieces.

We were able to relate all of the ingredients of the factorization formula except for $\vect{S}^{\rm unmeas}$ to analogous quantities that have previously been calculated in the context of $e^+ e^-$ collisions to NLL' accuracy. $\vect{S}^{\rm unmeas}$ was explicitly computed for the case of dijet production (for which all Wilson lines are coplanar) in terms of color operators $\vect{T}_i \cdot \vect{T}_j$ that encode the color correlations at this order. We in turn explicitly presented results for these color operators (which become matrices in color space) for the $qq' \to qq'$ channel, and plotted the corresponding distribution for the illustrative example where both jets are measured with $\tau_a$ for $a=0$ in the $\tau_a^1 = \tau_a^2$ bin. We also generalized the refactorization of  \Ref{Chien:2015cka} to include color-mixing effects and found that, as was already seen in $e^+e^-$, the normalization of the cross section and the corresponding scale uncertainty were reduced. Using the results of \Ref{Chien:2015cka}, our results can now be straightforwardly extended to NNLL for any combination of measured (at least for $a=0$) and unmeasured jets. The non-global logarithms which we do not include and would appear in a fixed order calculation  of the soft function beginning at $\cO(\as^2)$ have arguments of order $\pTc \cR^2/p_T^J\tau_a$ which for the peak region of the distribution (where we trust our calculation) is $\cO(1)$ to within a decade.

Armed with this foundation, we can now (after including all the partonic channels) make meaningful comparisons with Monte Carlo event generators and directly with data. It will be of particular interest to study the sensitivity of the proposed, factorized cross section to effects like multiple parton interactions. Other observables that are sensitive to radiation near the beam pipes like beam thrust \cite{Stewart:2010pd} have been noted to receive $\cO(1)$ corrections from these effects. We expect that our observables will be less sensitive to this effect because the jets are  isolated and the unmeasured beam functions should not be sensitive to radiation near the beam pipe. We also hope to be able to incorporate other effects with the recent developments for NGLs as discussed in the Introduction. In addition, the authors, together with other collaborators \cite{ee}, are actively involved in extending the results of this paper to cross sections for jets in which there is an identified heavy hadron. The work of Refs.~\cite{Procura:2009vm,Liu:2010ng,Jain:2011xz,Jain:2011iu,Procura:2011aq} shows that these cross sections  can be calculated by replacing the jet function for the jet with the identified hadron with so-called fragmenting jet functions. These are related to the well-known fragmentation functions by a matching calculation at the jet energy scale. These calculations  will be applied to the production of jets with open heavy flavor and heavy quarkonia, especially $J/\psi$ and $\Upsilon$. The cross sections will take essentially the same form as the cross sections in this paper, with an additional  convolution of the cross section with the heavy quark or quarkonium fragmentation 
 as well as a modified $f_J$ factor that depends on the matching coefficients in the fragmenting jet function. We expect to compare these predictions  to Monte Carlo event generators and LHC measurements \cite{ee}.


\acknowledgments
We would like to thank Christopher Lee, Daekyoung Kang, and Wouter Waalewijn for helpful discussions, and Christopher Lee for reviewing this manuscript. AH was supported by a Director's Fellowship from the  LANL/LDRD program and the DOE Office of Science under Contract DE-AC52-06NA25396. TM and YM are supported in part by the Director, Office of Science, Office of Nuclear Physics, of the U.S. Department of Energy under grant numbers DE-FG02-05ER41368. TM and YM also acknowledge the hospitality of the theory groups at Brookhaven National Laboratory, Los Alamos National Laboratory, Duke-Kunshan University, and UC-Irvine for their hospitality during the completion of this work.

\appendix

\section{Calculations of Soft Function Components}
\label{app:soft}

In this Appendix, we calculate the various components needed for $S^{\rm unmeas}$. As explained in the main body of the text, we only calculate combinations of terms that explicitly remove radiation out of the beams, i.e., with $y > \yc$ or $y<-\yc$. We use the definitions $c_J \equiv \vec n_J \cdot \vec n_B$, $s_J \equiv (1-c_J^2)^{1/2}$, $c_i \equiv \cos\theta_i$, and $s_i \equiv \sin\theta_i$. All the expressions are  special cases of the general form \eq{Iij-integral} in the planar limit, given by the substitution in \eq{planar}.   For subtraction terms $S_{ij}^k$ defined in \eq{Sijk} there is an additional factor of $-\Theta_\cR^k$ given in \eq{thetas}.
\subsection{Beam-Beam Interference Terms}
\label{app: bb}

We first calculate the beam-beam interference with the gluon out of the beams 
\begin{align}
\cI_{B\Bbar}^{\rm out} &\equiv \cI_{B\Bbar}^{\rm incl} +\cI_{B\Bbar}^B + \cI_{B\Bbar}^{\Bbar}  \nn\\
&= \frac{ e^{\gamma_E \epsilon}}{\sqrt \pi \Gamma(1/2-\epsilon)} \int_0^\pi\! d\theta_1 \sin\theta_1 \frac{1}{1-c_1}\frac{1}{1+c_1} \int_0^\pi d\theta_2 \sin^{-2\epsilon}\theta_2 \nn\\
&=\frac{e^{\gamma_E \epsilon}}{ \Gamma(1-\epsilon)} \int_{-\tanh \yc}^{\tanh\yc}\! \frac{dc_1}{1-c_1^2} \nn\\
&= \log \frac{1+ \tanh\yc}{1-\tanh\yc}   \nn\\
&= 2 \yc
\,.\end{align}

The region that must be added to remove radiation in the jets goes as $\cR^2$ and so is power suppressed for small jets, but we record it here for completeness. In a frame where the jet is perpendicular to the beam, 
\begin{align}
I_{B\Bbar}^J &= \frac{e^{\gamma_E \epsilon}}{\sqrt \pi \Gamma(1/2-\epsilon)} \int_0^R\! d\theta_1 \sin^{1-2\epsilon}\theta_1\int_0^\pi d\theta_2 \sin^{-2\epsilon}\theta_2 \big[1-(s_1 c_2)^2 \big]^{-1+ \epsilon}
\,.\end{align}
In this frame ($\theta_J = \pi/2$), we can make the substitution $R\to \cR \sin \pi/2 = \cR$ to get a frame invariant result. This gives
\begin{align}
\cI_{B\Bbar}^J &=    \frac{1}{2} \log(1-\cR^2) - \epsilon \bigg( \frac{\pi^2}{12} - \frac{1}{2} \Li_2(1-\cR^2) \bigg) = \cO(\cR^2)
\,.\end{align}

\subsection{Beam-Jet Interference Terms}
\label{app:bj}

The beam-jet interference term with the gluon out of both beams is simplest to compute in the polar coordinates about the beam axis. Defining $\cos \theta_c \equiv t_c \equiv \tanh \yc$, it can be written as
\begin{align}
\cI_{BJ}^{\rm out}  &\equiv \cI_{BJ}^{\rm incl} +\cI_{BJ}^B + \cI_{BJ}^{\Bbar} \nn\\
&=\frac{(1-c_J) e^{\gamma_E \epsilon}}{2 \sqrt \pi \Gamma(1/2-\epsilon)} \int_{\theta_c}^{\pi-\theta_c}\!\!\!d\theta_1\, \sin\theta_1 \int_0^\pi \!d\theta_2\, \sin^{-2\epsilon}\theta_2 \frac{1}{1-c_1} \frac{1}{1-c_J c_1 - s_J s_1 c_2} \nn\\
&= \frac{e^{\gamma_E \epsilon}}{2} \int_{-t_c}^{t_c} \!\frac{d c_1}{1-c_1^2} \frac{1-c_1}{1-c_1 c_J}  {}_{2}\tilde{F}_1(1/2,1; 1-\epsilon; z)
\,,\end{align}
where $z = (1-c_1^2)(1-c_J^2)/(1-c_1 c_J)^2$. We can proceed by extracting the $c_J = c_1$ singular via the identity
\begin{align}
\label{eq:2F1identity}
{}_{2}\tilde{F}_1\big(\frac{1}{2},1; 1-\epsilon; z\big) &= \frac{\sqrt \pi}{\Gamma(1/2-\epsilon)\cos \pi\epsilon} \bigg[ z^\epsilon\bigg(\frac{1-c_1 c_J}{\abs{c_1-c_J}}\bigg)^{\!1+2\epsilon}  \nn\\
& \qquad \qquad \qquad \qquad \qquad+ \frac{\epsilon \sqrt \pi}{\Gamma(1-\epsilon)} {}_{2}\tilde{F}_1\big( \frac{3}{2},1; \frac{3}{2}+\epsilon; 1-z\big) \bigg]
\,.\end{align}
The singularities are regulated by the $\abs{c_1-c_J}^{-1-2\epsilon}$ in the first term in brackets on the right hand side of \eq{2F1identity} (and the second term is finite and $\cO(\epsilon)$). After adding and subtracting the rest of the functional dependence on $c_1$, $f(c_1)$, at the point $c_1 = c_J$ (so that $\abs{c_1-c_J}^{-1-2\epsilon}(f(c_1) - f(c_J))$ can safely be expanded in $\epsilon$) and performing some algebra, we arrive at the result
\begin{align}
\cI_{BJ}^{\rm out}  &= \frac{e^{\gamma_E \epsilon}}{\Gamma(1-\epsilon)}\bigg\{ -\frac{1}{2\epsilon} + \frac{1}{2}\bigg[ \log\big(e^{2(\yc - y_J)}-1 )+ \log \big(1-e^{-2(\yc + y_J} \big)\bigg] \nn\\
& \quad \, -\epsilon \bigg[ \frac{1}{2}\log^2\big(1-e^{-2(\yc - y_J)}\big)+\Li_2\big(e^{-2(\yc - y_J)}\big) + \frac{1}{2}\log^2\big(1-e^{-2(\yc + y_J)}\big)\bigg]\bigg\} \nn\\
&=  \frac{e^{\gamma_E \epsilon}}{\Gamma(1-\epsilon)}\bigg[ -\frac{1}{2\epsilon} + \yc - y_J + \cO(e^{-\yc}) \bigg]
\,.\end{align}

For the jet region subtraction term $S_{JB}^J$, in coordinates about the jet axis, we have
\begin{align}
\cI_{BJ}^{J }  &= \frac{(1-c_J) e^{\gamma_E \epsilon}}{2 \sqrt \pi \Gamma(1/2-\epsilon)} \int_{0}^{R}\!\!\!d\theta_1\, \sin^{1-2\epsilon}\theta_1 \int_0^\pi \!d\theta_2\, \sin^{-2\epsilon}\theta_2 \nn\\
& \qquad \times\frac{1}{1-c_1} \frac{1}{1-c_J c_1 - s_J s_1 c_2} \Big[ 1-(c_J c_1 + s_J s_1 c_2)^2 \Big]^{\epsilon} \nn\\
&= \frac{(1-c_J)e^{\gamma_E \epsilon}}{2 \sqrt \pi \Gamma(1/2-\epsilon)} \int_{\cos R}^1 \! dc_1 (1-c_1)^{-1-\epsilon} f(c_1)
\,,\end{align}
where we defined
\begin{align}
f(c) &= (1+c)^{-\epsilon} \int_0^\pi\! d \theta_2 \sin^{-2\epsilon}\theta_2 \frac{\big[1-(c_J c + s_J (1-c^2)^{1/2} c_2)^2 \big]^\epsilon}{1-c_J c - s_J (1-c^2)^{1/2}  c_2 }
\,.\end{align}
Up to corrections that scale as $\cO(\cR^2)$, we can set $f(c) = f(1)$  which is just
\begin{align}
f(1) = \frac{2^{-\epsilon}}{1-c_J}s_J^{2\epsilon}  \frac{\sqrt \pi  \Gamma(1/2-\epsilon)}{\Gamma(1-\epsilon)} 
\,.\end{align} 
Using the substitution \eq{rescale}, we find
\begin{align}
\cI_{BJ}^{J}&= \frac{e^{\gamma_E \epsilon}}{\Gamma(1-\epsilon)} \frac{1}{2\epsilon} \cR^{-2\epsilon} + \cO(\cR^2) 
\,.\end{align}

\subsection{Jet-Jet Interference Terms}
\label{app:jj}

For the jet-jet interference terms, we work in coordinates about the jet axes in the frame where they are back-to-back, and then convert to lab frame variables. For the term with the gluon allowed anywhere, labeling the jets as $1$ and $2$, we have
in the frame of back-to-back jets, 
\begin{align}
\cI_{BJ}^{J }  &= \frac{e^{\gamma_E \epsilon}}{ \sqrt \pi \Gamma(1/2-\epsilon)} \int_{0}^{R}\!\!\!d\theta_1\, \sin^{1-2\epsilon}\theta_1 \int_0^\pi \!d\theta_2\, \sin^{-2\epsilon}\theta_2 \nn\\
& \qquad \times\frac{1}{1-c_1} \frac{1}{1+c_1} \Big[ 1-(c_J c_1 + s_J s_1 c_2)^2 \Big]^{\epsilon} \nn\\
&= \frac{e^{\gamma_E \epsilon}}{ \sqrt \pi \Gamma(1/2-\epsilon)} \,2\! \int_{0}^1 \! dc_1 (1-c_1)^{-1-\epsilon} g(c_1)
\,,\end{align}
where we defined
\begin{align}
g(c) &= (1+c)^{-1-\epsilon} \int_0^\pi\! d \theta_2 \sin^{-2\epsilon}\theta_2 \big[1-(c_J c + s_J (1-c^2)^{1/2} c_2)^2 \big]^\epsilon
\,.\end{align}
As before, we can add and subtract $g(1)$, with
\begin{align}
g(1) = \frac{2^{-1-\epsilon}}{1-c_J}s_J^{2\epsilon}  \frac{\sqrt \pi  \Gamma(1/2-\epsilon)}{\Gamma(1-\epsilon)} 
\,,\end{align} 
and expand the part of the integrand with $(1-u)^{-1-\epsilon} (f(u) - f(1))$ in $\epsilon$. To evaluate the result,  note that
\begin{align}
\label{eq:hdef}
h(c_J, c_1) &\equiv \frac{1}{\pi} \int_0^\pi \!d\theta \log\frac{1-(c_1 c_J+ (1-c_1^2)^{1/2}(1-c_J^2)^{1/2} \cos \theta)^2}{1-c_J^2} \nn\\
&= 
\begin{cases}
\ln \Big[\frac{1-c_1^2}{1-c_J^2} \Big(\frac{1+\abs{c_J}}{2}\Big)^{\!2}\Big] &\mbox{for } \abs{c_1} < \abs{c_J}
\vspace{.5em} \\
2 \ln \frac{1+\abs{c_1}}{2} &\mbox{for } \abs{c_1} > \abs{c_J}
\end{cases} 
\,,\end{align}
and that
\begin{align}
\int_0^1\! \frac{d c_1}{1-c_1^2} f(c_J, c_1) = -\frac{\pi^2}{6}+\frac{1}{2}\log^2 \frac{1-c_J}{1+c_J} \, ,
\end{align}
to finally obtain
\begin{align}
\cI_{12}^{\rm incl} &=  \frac{e^{\gamma_E \epsilon}}{ \Gamma(1-\epsilon)}\bigg(\frac{1-\cos^2\theta_J}{4}\bigg)^{\!\epsilon} \bigg[- \frac{1}{\epsilon} + \frac{\epsilon}{2}\log^2 \frac{1-c_J}{1+c_J} \bigg]
\,.\end{align}
Noting that $c_J \equiv \cos \theta_J$ in the back-to-back frame is related to the jet rapidities in the lab frame via $\cos \theta_J = \tanh \Delta y/2$ (cf. \eq{b2bframeangle}), we find
\begin{align}
\cI_{12}^{\rm incl}  &= -  \frac{e^{\gamma_E \epsilon}}{ \Gamma(1-\epsilon)} \big(2\cosh (\Delta y/2)\big)^{-2\epsilon}\bigg[\frac{1}{\epsilon} - \frac{\epsilon}{2} (\Delta y)^2 \bigg] \nn\\
&=  - \big(2\cosh (\Delta y/2)\big)^{-2\epsilon}\bigg[\frac{1}{\epsilon} - \frac{\epsilon}{2} (\Delta y)^2 - \frac{\pi^2}{12} \bigg] 
\,.\end{align}

For the jet region subtraction terms, we have 
\begin{align}
 \cI_{12}^{1} &= \frac{e^{\gamma_E \epsilon}}{ \sqrt \pi \Gamma(1/2-\epsilon)} \,2\! \int_{\cos R}^1 \! dc_1 (1-c_1)^{-1-\epsilon} g(c_1)
\,,\end{align}
which now involves the integral of $h(c_J, c_1)$ (cf. \eq{hdef}) over the range $c_1 \in (\cos R, 1)$ with $c_J <  \cos R$ (so only the case $\abs{c_1} > \abs{c_J}$ is needed). After some algebra and using the substitution $\tan R/2 \to \cR/(2\cosh\Delta y/2)$, we arrive at the result
\begin{align}
\cI_{12}^{1} &=   \frac{e^{\gamma_E \epsilon}}{ \Gamma(1-\epsilon)}  \frac{1}{2\epsilon} \cR^{-2\epsilon}
\,.\end{align}

\section{Review of Renormalization and RG Evolution}
\label{app:RGE}

In this Appendix we review renormalization and RG evolution for multiplicatively renormalized functions that are trivial in color-space (namely, the unmeasured jet and beam functions) and for functions of $\tau_a$ which renormalize and evolve via a convolution (such as measured jet functions and the measured part of the soft function). The RGE for the non-trivial color-space matrix components of the hard and (unmeasured) soft functions is derived explicitly in  \ssec{hardRGE} and \ssec{softRGE}, respectively.

Renormalization of the multiplicative-type functions which are trivial in color-space takes the form
\be
 F^{\text{bare}} = Z_F(\mu)F(\mu)\,.
 \ee
The independence of the left-hand side on $\mu$ gives rise the RG evolution equation,
\be
\label{eq:FRGE}
\mu \frac{d}{d\mu}F(\mu) =  \gamma_F(\mu) F(\mu) \,,\ee
where the anomalous dimension $\gamma_F$ is defined as
\be
\label{eq:anom-from-Z-mult}
\gamma_F(\mu) = - \frac{1}{Z_F(\mu)}\mu \frac{d}{d \mu} Z_F(\mu)
\,,\ee
and to all orders in $\alpha$ takes the form,
\begin{equation}
\label{eq:gammaF}
\gamma_F(\mu) = \Gamma_F[\alpha] \ln\frac{\mu^2}{m_F^2} + \gamma_F[\alpha]\,.
\end{equation}
where $\Gamma_F[\alpha]$ and $\gamma_F[\alpha]$ have the expansions
\begin{align}
\label{eq:Gammaexpansion}
\Gamma_{\!F}[\as] = \left( \frac{\as}{4\pi}\right) \Gamma_{\!F}^0 + \left( \frac{\as}{4\pi}\right)^2 \Gamma_{\!F}^1 + \cdots
\end{align}
and
\begin{align}
\label{eq:gammaexpansion}
\gamma_F[\as] = \left( \frac{\as}{4\pi}\right) \gamma_{F}^0 + \left( \frac{\as}{ 4\pi}\right)^2 \gamma_{F}^1 + \cdots
\,.\end{align}
The RGE \eq{FRGE} has the solution
\begin{align}
F(\mu)
& = \Pi_F(\mu,\mu_0)F(\mu_0) \,,
\end{align}
where the evolution kernel $\Pi_F$ is given by
\begin{equation}
\label{eq:kernel}
\Pi_F(\mu,\mu_0)= e^{K_F(\mu,\mu_0)} \left(\frac{\mu_0}{m_F}\right)^{\omega_F(\mu,\mu_0)}\,,
\end{equation}
where $K_F(\mu,\mu_0)$ and $\omega_F(\mu,\mu_0)$ will be defined below in \eq{kernelparams}.

Renormalization of functions which depend on the jet shape, $\tau_a$, 
takes the form of a convolution,
\begin{equation}
F^{\text{bare}}(\tau_a) = \int d\tau_a' Z_F(\tau_a - \tau_a',\mu)F(\tau_a',\mu)\,,
\end{equation}
and satisfies the RGE
\begin{align}
   \label{eq:RGEtau}
  & \mu \frac{d}{d\mu} F(\tau_a, \mu)= \int_{}^{}d \tau_a' \, \gamma_F (\tau_a-\tau_a', \mu) F(\tau_a', \mu)
\,,\end{align}
with the anomalous dimension in this case given by
\be
\label{eq:anom-from-Z-convolve}
\gamma_F(\tau_a, \mu) = - \int\! d\tau_a'\, Z_F^{-1} (\tau_a - \tau_a', \mu) \, \mu \frac{d}{d \mu} Z_F(\tau_a', \mu)
\,,\ee
and taking the general form
\begin{align}
\label{eq:gammaFtau}
\gamma_F (\tau_a, \mu) = -  \Gamma_F [\as] \left( \frac{2}{j_F} \left[ \frac{\Theta(\tau_a)}{\tau_a}\right]_{\plus} -\ln \frac{\mu^2}{m_F^2} \,\delta(\tau_a)\right)  + \gamma_F[\as] \delta(\tau_a)
\,.\end{align}
The solution of \eq{RGEtau} is
\begin{align}
F(\tau_a, \mu) = \int\!d \tau' \, U_F(\tau_a-\tau_a', \mu, \mu_0) F(\tau_a', \mu_0)
\label{eq:Fconvol}
\,,\end{align}
where to all orders in $\alpha_s$  the evolution kernel $U_F$ is given  by  \cite{Korchemsky:1993uz,Becher:2006mr,Balzereit:1998yf,Neubert:2005nt,Fleming:2007xt}
\begin{align}
 U_F(\tau_a, \mu, \mu_0)= \frac{e^{{K}_F + \gamma_E{\omega}_F}}{\Gamma(-{\omega}_F)} \left(\frac{\mu_0}{m_F}\right)^{j_F\omega_F} \left[\frac{\Theta(\tau_a)}{(\tau_a)^{1+{\omega}_F}}\right]_{\plus}
 \label{eq:kernelF}
\,,\end{align}
where $\gamma_E$ is the Euler constant.

The exponents ${\omega}_F(\mu,\mu_0)$ and ${K}_F(\mu,\mu_0)$ of \eqs{kernel}{kernelF} are given by (where we set $j_F = 1$ in the multiplicative case of \eq{gammaF})
\begin{subequations}
   \label{eq:kernelparams}
\begin{align}
\label{eq:omegaF}
  {\omega}_F(\mu,\mu_0) & \equiv \frac{2}{j_F}\int_{\as(\mu_0)}^{\as(\mu)}\frac{d\alpha}{\beta[\alpha]} \Gamma_{\!F}[\alpha] \,, \\
   \label{KF}
   {K}_F(\mu,\mu_0)& \equiv \int_{\as(\mu_0)}^{\as(\mu)}\frac{d\alpha}{\beta[\alpha]} \gamma_F[\alpha]+2\int_{\as(\mu_0)}^{\as(\mu)}\frac{d\alpha}{\beta[\alpha]} \Gamma_{\!F}[\alpha]\int_{\as(\mu_0)}^{\alpha}\frac{d \alpha'}{\beta[{\alpha'}]}
\,.\end{align}
\end{subequations}
At NLL (and NLL') accuracy we can write $\omega_F(\mu,\mu_0)$ and $K_F(\mu,\mu_0)$ as
\begin{subequations}
  \label{eq:kernelparamsNLL}
\begin{align}
\label{eq:omegaFNLL}
 \omega_F(\mu, \mu_0) \Big\vert_{\rm NLL}  &=-\frac{\Gamma_{\!F}^0}{j_F\, \beta_0} \left[\ln{r}+\left(\frac{\Gamma_{\cusp}^1}{\Gamma_{\cusp}^0}-\frac{\beta_1}{\beta_0}\right)\frac{\as(\mu_0)}{4\pi}(r-1)\right] \,,\\
  \label{KFNLL}
K_F(\mu_,\mu_0) \Big\vert_{\rm NLL} &=-\frac{\gamma_{F}^0}{2\beta_0}\ln {r} - \frac{2\pi\Gamma_{\!F}^0}{(\beta_0)^2}\bigg[\frac{r-1-r\ln{r}}{\as(\mu)} \nn\\
  & \qquad \qquad   +\left(\frac{\Gamma^1_{\cusp}}{\Gamma^0_{\cusp}}-\frac{\beta_1}{\beta_0}\right)\frac{1-r+\ln{r}}{4\pi}+\frac{\beta_1}{8\pi\beta_0}\ln^2{r}\bigg]
\,,\end{align}
\end{subequations}
where $r =\as(\mu)/\as(\mu_0)$, which can be evaluated at two loops via the equation,
\begin{equation}
\label{eq:2loopalphas}
\frac{1}{\as(\mu)} = \frac{1}{\as(M_Z)} + \frac{\beta_0}{2\pi}\ln\left(\frac{\mu}{M_Z}\right) + \frac{\beta_1}{4\pi\beta_0}\ln\left[1+\frac{\beta_0}{2\pi}\as(M_Z)\ln\left(\frac{\mu}{M_Z}\right)\right]
\,,\end{equation}
with $\beta_0, \beta_1$  are the one-loop and two-loop  coefficients of the beta function,
\begin{equation}
\beta[\alpha_s] = \mu\frac{d \alpha_s}{d \mu} = -2\alpha_s\left[\beta_0\left(\frac{\alpha_s}{4\pi}\right) + \beta_1\left(\frac{\alpha_s}{4\pi}\right) ^2 +\cdots\right]\,,
\end{equation}
and where (with $\TR$ set to $1/2$)
\begin{align}
\label{eq:beta-pieces}
\beta_0=\frac{11C_A}{3} - \frac{2\NF}{3} \qquad {\rm and } \qquad \beta_1= \frac{34C_A^2}{3}-\frac{10C_A \NF}{3} -2\CF \NF
\,.\end{align}

In \eq{kernelparamsNLL}, we have used that  $\Gamma_F[\alpha_s]$ for $F=H,J,S$ (hard, jet, and soft) is proportional to $\Gamma_{\rm cusp}[\alpha_s]$, where
 \begin{align}
 \label{eq:gamma-gammacusp}
 & \Gamma_{\rm cusp}[\as]=\left(\frac{\as}{4\pi}\right)\Gamma^0_{\cusp}+\left(\frac{\as}{4\pi}\right)^2 \Gamma^1_{\cusp}+\cdots
 \,.\end{align}
Here $\Gamma_\cusp^0=4$ and the ratio of the one-loop and two-loop coefficients of $\Gamma_{\rm cusp}$ is \cite{Korchemsky:1987wg}
 \begin{align}
 \label{eq:Gammac10}
 &\frac{\Gamma_{\cusp}^1}{\Gamma^0_{\cusp}}=\left(\frac{67}{9}-\frac{\pi^2}{3}\right)C_A-\frac{10 \NF}{9}
\,.\end{align}
At NLL', we will need both $\Gamma_{\cusp}^1$ and $\beta_1$ in the expressions of $\omega_F$ and $K_F$ for  NLL' resummation.

\bibliography{shapes}

\end{document}